\documentclass[onecolumn,amsmath,amssymb,aps,superscriptaddress,nofootinbib]{revtex4-2}
\usepackage{dcolumn}                                   

\usepackage{graphicx,float,wrapfig}
\usepackage{hyperref}
\usepackage{bm}         
\usepackage{amssymb}  
\usepackage{babel}
\usepackage{comment}
\usepackage{verbatim}
\usepackage{amsmath}
\usepackage[export]{adjustbox}
\expandafter\def\csname ver@subfig.sty\endcsname{}
\usepackage{amsthm}
\usepackage{subfig}
\usepackage{subcaption}
\usepackage{mathtools}
\theoremstyle{remark}
\usepackage{xcolor}

\newtheorem*{Pre*}{Proof}

\usepackage{braket}
\usepackage{ulem} 

\usepackage{float}
\usepackage{aliascnt}
\usepackage{afterpage}
\newaliascnt{eqfloat}{equation}
\newfloat{eqfloat}{h}{eqflts}
\floatname{eqfloat}{Equation}
\usepackage{tikz}
\usetikzlibrary{quantikz}
\usepackage{leftindex}
\usepackage{mathdots}

\newtheorem{theorem}{Theorem}[section]

\newtheorem{lemma}[theorem]{Lemma}

\begin{document}

\title{Trotter-based quantum algorithm for solving transport equations with exponentially fewer time-steps}

\author{Julien Zylberman}
\affiliation{Sorbonne Université, Observatoire de Paris, Université PSL, CNRS, LUX, F-75005 Paris, France}

\author{Thibault Fredon}
\affiliation{Plasma Science and Fusion Center, Massachusetts Institute of Technology, Cambridge, Massachusetts 02139, USA}

\author{Nuno F. Loureiro}
\affiliation{Plasma Science and Fusion Center, Massachusetts Institute of Technology, Cambridge, Massachusetts 02139, USA}
\author{Fabrice Debbasch}
\affiliation{Sorbonne Université, Observatoire de Paris, Université PSL, CNRS, LUX, F-75005 Paris, France}

\begin{abstract}

The extent to which quantum computers can simulate physical phenomena and solve the partial differential equations (PDEs) that govern them remains a central open question. In this work, one of the most fundamental PDEs is addressed: the multidimensional transport equation with space- and time-dependent coefficients. We present a quantum numerical scheme based on three steps: quantum state preparation, evolution, and measurement of relevant observables. The evolution step combines a high-order centered finite difference with a time-splitting scheme based on product formula approximations, also known as Trotterization. We introduce novel vector-norm analysis and prove that the number of time-steps can be reduced by a factor exponential in the number of qubits compared to previously established operator-norm analysis, thereby significantly lowering the projected computational resources. We also present efficient quantum circuits and numerical simulations that confirm the predicted vector-norm scaling. We report results on real quantum hardware for the one-dimensional convection equation, and solve a non-linear ordinary differential equation via its associated Liouville equation, a particular case of transport equations. This work provides a practical framework for efficiently simulating transport phenomena on quantum computers, with potential applications in plasma physics, molecular gas dynamics and non-linear dynamical systems, including chaotic systems.
\end{abstract}

\maketitle

\tableofcontents
\section*{Introduction}

In recent years, experimental advances have led to the first Noisy Intermediate Scale Quantum (NISQ) computers made of hundreds of physical qubits, and experimental evidence of error-corrected qubits suggests that quantum computers capable of outperforming classical ones may soon become available \cite{marques2022logical,krinner2022realizing,sundaresan2023demonstrating,paetznick2024demonstration,google2023suppressing,google2025quantum,rousseau2025enhancing}. However, realizing a quantum advantage depends not only on hardware progress but also on the development of more efficient quantum algorithms.

Solving partial differential equations (PDEs) is among the most computationally demanding tasks in classical computing. Many PDEs are used to model complex phenomena that are nonlinear, multidimensional, or multiscale. Capturing small-scale dynamics or accurately simulating long-time evolution requires substantial computational resources, often relying on High-Performance Computing\footnote{For instance, the 2023 NERSC Annual report estimates that $14\%$ of the $57$ millions core-hours have been allocated to Fusion and plasma simulations (see https://www.nersc.gov/about/annual-reports), and the DOE INCITE 2025 projects allocated $40$ millions core-hours over 81 projects, of which
over $11\%$ focused on plasma and fluid dynamics problems (see https://doeleadershipcomputing.org/awardees/).}. Quantum computing offers a potential alternative for solving PDEs, with possible advantages in memory efficiency and computational speed arising from quantum superposition and entanglement. Although it remains unclear whether a genuine speedup will be achieved, the development of novel quantum numerical schemes represents a promising research direction for advancing computational treatment of PDEs.

For this purpose, many quantum approaches have been developed to solve PDEs. For linear PDEs, the different methods can be classified into three main categories. First, the methods based on the Quantum Linear Solver Algorithms (QLSA), for which the initial PDE problem is converted into a linear system of equations $Ax=b$ that requires to be inverted \cite{HHL09,childs2017quantum,berry2014high,berry2017quantum,childs2021high,Krovi2023,morales2024quantum}. Second, the methods that transform the initial PDE into a set of ordinary differential equations (ODEs) $df/dt=Af$ that are solved by implementing the associated integrator using product formula methods \cite{bosse2024efficient}, linear combination of unitaries \cite{an2023linear,an2023quantum,an2024laplace} or quantum signal processing protocols \cite{low2017optimal,motlagh2024generalized}. Note that the underlying algorithm strongly depends on the properties of the operator $A$: in particular, whether 
$A$ is time-dependent or time-independent, and whether it is anti-Hermitian, in which case Hamiltonian simulation methods apply \cite{morales2024quantum,berry2020time, An2021,an2022time,watkins2024time,berry2015hamiltonian,berry2015simulating,berry2024doubling,childs2012hamiltonian,an2023linear,an2023quantum,an2024laplace,mangin2024efficient}. Lastly, optimization-based methods such as variational quantum algorithms, quantum machine learning, or physics-informed quantum neural networks can address the problem of solving differential equations \cite{liu2021variational,cerezo2021variational,sugaya2025variational,setty2025self}. 

For non-linear PDEs, the intrinsic linearity of digital quantum computers limits the development of efficient and accurate quantum numerical schemes. To overcome this issue, a few approaches have been proposed to transform the non-linear problem into a linear one. The Carleman linearization method --- which transforms a set of non-linear ODEs into an infinite set of linear ODEs that requires truncation \cite{Liu2021,Krovi2023} --- has been recently applied to solve various PDEs \cite{liu2023efficient,sanavio2024lattice,li2025potential}. Similarly, homotopy perturbation methods recast non-linear ODEs into an infinite system of linear ODEs \cite{xue2021quantum,xue2022quantum}. In some particular cases, non-linear mappings allow to transform a non-linear PDE problem into a linear one. For instance, the Madelung transform maps the non-linear inviscid Burger equation into a linear Schrödinger-type equation that can be solved using quantum algorithms for linear PDEs \cite{ZDMD22,meng2023quantum,meng2024simulating}. Quantization methods provide another mapping of classical non-linear dynamics into an infinite dimensional Fock-space in which the dynamics is linear \cite{shi2024simulating,may2024nonlinear,may2025second}. The Koopman-Von Neumann (KVN) approach embeds non-linear dynamics into a linear equation expressed in the associated phase-space, enabling the simulation of the dynamics on quantum computers \cite{joseph2020koopman,novikau2025quantum}. In particular, when the dynamics is Hamiltonian, the associated phase-space equation has a unitary evolution, and is often referred to as the Liouville equation \cite{joseph2020koopman}. Variational quantum algorithms, quantum machine learning, and physics-informed quantum neural networks also address non-linear problems \cite{setty2025self,lubasch2020variational,kyriienko2021solving,mouton2024deep,sarma2024quantum}.

In this work, we focus on one of the most fundamental PDEs: the transport equation\footnote{The transport equation is also known as the convection equation, or the advection equation. In this work, we refer to the transport equation in the sense of a pure advection equation, i.e., without diffusion or source terms.}. Transport phenomena are ubiquitous in physics. For instance, the collisionless Boltzmann equation is a type of transport equation which describes the kinetic transport of particles in external force fields, with applications in plasma physics, molecular gas dynamics, and astrophysics \cite{vincenti1966introduction,cercignani1988boltzmann,aoki1985hydrodynamics,hersant2009pressure,evans2011discontinuous}. Transport equations can also be used to simulate non-linear Hamiltonian dynamics via the associated Liouville equation, opening the possibility of studying chaotic systems, resonances, and instabilities \cite{hirsch2013differential,arnold2006mathematical}. 

In the literature, a few works have already addressed transport equations in one dimension or considered constant position-independent coefficients \cite{sato2024hamiltonian,hu2024quantum,brearley2024quantum,jin2023time}. In the following, we present a more general approach to solve multidimensional and anisotropic transport equations with space- and time-dependent coefficients. We introduce a quantum numerical scheme composed of three steps: an initialization to load the initial condition into a qubit state; an evolution step based on high-order finite difference and product formula approximations, also known as Trotterization; and a measurement step to estimate quantities of interest. A vector norm analysis is provided for the space and time discretizations with novel bounds on the product formula approximations. These bounds improve upon previous results based on operator-norm analysis by a factor of $\Theta(4^n)$, where $n$ is the number of qubits per dimension. As a result, the number of time-steps required to guarantee an $\epsilon$-approximation of the solution at time $T$ is reduced exponentially in $n$, from $O(4^n T^2/\epsilon)$ to $O(T^2/\epsilon)$, significantly improving the overall complexity of the quantum numerical scheme. Numerical simulations of transport equations and non-linear Hamiltonian dynamics illustrate the different results, showcasing the approximations and quantum circuit complexity scalings.

In section \ref{sec: problem and approach}, the PDE problem, the high-order finite differences, the qubit encoding and the product formula are formulated. In section \ref{sec: quantum numerical scheme}, the quantum numerical scheme is presented with technical details about the different steps. In section \ref{sec : error analysis}, an error and complexity analysis is introduced. Generic error bounds based on operator norm are recalled and the novel vector-norm analysis is introduced. In section \ref{sec: application}, we present simulations of the collisionless Boltzmann equation that illustrate the theoretical scalings, along with quantum-hardware implementations of the one-dimensional convection equation and simulations of a non-linear Hamiltonian dynamical system. Finally, the results are discussed in section \ref{sec: discussion}.

\section{Problem and approach}
\label{sec: problem and approach}
\subsection{Problem formulation}
\label{sec:problem_formulation}
Let $\vec{c}=\vec{c}(x,t)$ be a time dependent vector field defined on $\mathbb{R}^d\times[0,T]\rightarrow\mathbb{R}^d$ for some time $T>0$. Consider the following multidimensional transport equation with $ \vec{x}\in \mathbb{R}^d, t\in[0,T]$: 
\begin{equation}
\begin{split}
    &\partial_t f +\vec{c}\cdot\vec{\nabla}f=0, \\
    &f(\vec{x},t=0)=f_0(\vec{x}), \text{  } 
\label{Transport equation}
\end{split}
\end{equation}

where $f_0:\mathbb{R}^d\rightarrow\mathbb{R}$ is a given initial condition with compact support and $f:\mathbb{R}^d\times[0,T]\rightarrow\mathbb{R}$ is the unknown. We assume that $\vec{c}$
is bounded, which implies that the solution $f$ propagates at a finite speed and remains supported within a bounded domain $\Omega\subset\mathbb{R}^d$ for the duration of the simulation. Therefore, we assume that $f$, $f_0$, and $\vec{c}$ are defined on $[0,1]^d$ and impose periodic boundary conditions. These conditions could be applied to simulate inherently periodic phenomena.

Since the components $c_j$ of the vector field $\vec{c}=(c_1,...c_d)$ are position- and time-dependent functions, the convergence, stability and efficiency of a numerical scheme directly depend on their regularity. In the following, we consider the case where $f_0$ and the $c_j$ functions are $(2p
+1)$-differentiable, ensuring the existence of a $(2p+1)$-differentiable solution $f$ and the convergence of high-order finite difference as proven in section \ref{subsec: space discretization error}. Additionally, we assume each function $c_j$ does not depend on the $j$-th variable $x_j$:
\begin{equation}
\forall j\in\{1,\hdots,d\},\text{  } \partial_{x_j}c_j=0,
\label{constraint}
\end{equation}
implying that $\vec{c}$ is divergence free. Under this last assumption, the solution function has a unitary evolution, opening the possibility of developing a numerical scheme that is naturally implementable on quantum computers. The consequences of removing this hypothesis are discussed in section \ref{sec: discussion}.

As detailed in section \ref{sec: application}, many applications are captured by these assumptions, such as the collisionless Boltzmann equation with external forces and the solution of non-linear Hamiltonian ODEs by considering their associated Liouville equations.

\subsection{Space discretization}
The space $[0,1]^d$ is uniformly discretized by introducing the discrete position variables $X_j\in \{0/N_j,1/N_j,...,(N_j-1)/N_j\}$, with $N_j=2^{n_j}$ the number of points on the $j^{\text{th}}$ axis. The discretization step associated with the $j^{\text{th}}$ variable is $\Delta x_j=1/N_j$. Derivatives are approximated using a central finite difference scheme of order $2p$:
\begin{equation}
     \partial_{x_j}f(\vec{x},t) \rightarrow  \frac{\sum_{k=-p}^p a_kf(\vec{x}+k\vec{x}_j,t)}{\Delta x_j}+O(\Delta x_j^{2p}),
\label{CFD order 2p}
\end{equation}

where $\vec{x}_j=(0,...,0,\Delta x_j,0,...0)$ is the vector adding a step $\Delta x_j$ with respect to the $j^{\text{th}}$ axis. The finite difference coefficients $a_k$ are solution of the system of equations $\sum_{k=-p}^p a_k k^j=\delta_{j,1}$, where $j \in \{0,...,2p\}$. Explicitly, one has $a_0=0$ and $a_k=(-1)^{k+1}(p!)^2/(k(p-k)!(p+k)!)$ for $k\in \{-p,-p+1,...p-1,p\}\backslash \{0\}$.

\subsection{Real space encoding}
The real space encoding consists in encoding the solution function into a $n=n_1+...+n_d$ qubit state:
\begin{equation}
    \ket{f}_t=\sum_{\vec{X} }f(\vec{X},t)\ket{\vec{X}} \in \mathcal{H},
\label{real space encoding}
\end{equation}
where $\mathcal{H}=\mathcal{H}_{x_1}\otimes...\otimes\mathcal{H}_{x_d}$ is the Hilbert space associated with the $n$ entangled qubits and each $\mathcal{H}_{x_j}$ is a $2^{n_j}$-dimensional Hilbert space. The ket vector $\ket{\vec{X}}$ is defined as the tensor product of the ket vector of each space $\ket{\vec{X}}=\ket{X_1}\otimes...\otimes\ket{X_d}$, where the discrete position $X_j$ is encoded through its dyadic expansion $\ket{X_j}=\ket{q_0^j}\otimes...\otimes\ket{q_{n_j-1}^j}$ with $X_j=\sum_{k=0}^{n_j-1}q_k^j/2^{k+1}\in  \{0/N_j,...,(N_j-1)/N_j\}$ and $ q_k^j\in\{0,1\}$.

The central finite difference derivative operator along the $j$-th axis is defined using the shift operators $\hat{S}_j =\sum_{X_j}\ket{X_j+\Delta x_j}\bra{X_j}$ as: 
\begin{equation}
    \hat{D}_{j}=\hat{I}_1\otimes\cdots\otimes \hat{I}_{j-1} \otimes -i\frac{\sum_{k=-p}^p a_k (\hat{S}_j)^k}{\Delta x_j}\otimes \hat{I}_{j+1} \otimes\cdots \otimes \hat{I}_d,
\label{2p discrete derivative operator}
\end{equation}
where the extra factor $i$ (with $i^2=-1$)  has been added in order to obtain an hermitian operator $\hat{D}_{j}^\dagger= \hat{D}_{j}$. 

Then, we define $\hat{\vec{X}}=(\hat{X}_1,...\hat{X}_d)$ where $\hat{X}_j=\hat{I}_1\otimes \cdots \otimes \hat{I}_{j-1}\otimes \sum_{X_j}X_j\ket{X_j}\bra{X_j}\otimes \hat{I}_{j+1}\otimes \cdots\otimes \hat{I}_d$ is the position operator associated with the $j$-th axis, and $\hat{c}_j(t)=c_j(\hat{\vec{X}},t)$ the diagonal operator with eigenvalues $c_j(\vec{X},t)$ and eigenvectors $\ket{\vec{X}}$. The multidimensional transport equation Eq.(\ref{Transport equation}) can now be recast as an ordinary differential equation:
\begin{equation}
\begin{split}
    \frac{d\ket{\tilde{f}}_t}{dt} &=-i(\sum_{j=1}^d \hat{c}_j(t)\hat{D}_j)\ket{\tilde{f}}_t, \\
    \ket{\tilde{f}}_{t=0}&=\ket{f_0},
\end{split}
\label{ODE}
\end{equation}
where $ \ket{f_0}=\frac{1}{||f_0||_{2,N}} \sum_{\vec{X}}f_0(\vec{X})\ket{\vec{X}} $ and $||f_0||_{2,N}=\sqrt{ \sum_{\vec{X}}|f_0(\vec{X})|^2}$ is a normalization factor \footnote{We keep track of the dependence of the norms with the number of grid points $N=N_1\times\hdots\times N_d$ to emphasize the difference with the $L^2$-norm of functions that is also used later.}. To be explicit, we distinguish  $\ket{\tilde{f}}_t$, the solution of the ODE problem, to the real space encoding $\ket{f}_t$ of the solution $f$ of the transport equation Eq.(\ref{Transport equation}).

\subsection{Time evolution with product formula}

First, notice that $\ket{\tilde{f}}_t$ has a unitary evolution: the central finite difference has been chosen to preserve the unitary evolution of the solution and assumption \ref{constraint} implies the commutation of the operators $\hat{c}_j(t)$ and $\hat{D}_j$: $\hat{c}_j(t)\hat{D}_j=\hat{D}_j\hat{c}_j(t)$. Consequently, the operator $\sum_{j=1}^d \hat{c}_j(t)\hat{D}_j$ is Hermitian and the evolution operator associated with the time evolution of $\ket{\tilde{f}}_t$ is unitary. The time dependence of the $c_j$ functions implies that the evolution of $\ket{\tilde{f}}_t$ is given by a time-ordered exponential:
\begin{equation}
    \hat{U}(T)=\hat{\mathcal{T}}e^{-i\int_0^T \big (\sum_{j=1}^d  \hat{c}_j(s)\hat{D}_j \big )ds},
\end{equation}
where $\hat{\mathcal{T}}$ is the time-ordering operator. In general, it is not possible to implement directly time-ordered exponentials on computers, whether classical or quantum. To overcome this issue, we consider the first order product formula, also called Trotterization, to decompose the time-ordered evolution operator into efficiently implementable unitary operators. The standard product formula denoted by the index $s$, and the generalized one denoted by $g$ are defined as \cite{An2021}:

\begin{equation}
\begin{split}
    \tilde{U}_{s}(t+h,t)&=\overleftarrow{\prod_{j=1}^{d}} \exp(-ih \hat{c}_j(t+h)\hat{D}_j),\\
    \tilde{U}_{g}(t+h,t)&=\overleftarrow{\prod_{j=1}^{d}} \exp(-i\int_{t}^{t+h}\hat{c}_j(s)ds\hat{D}_j),
\label{product formula}
\end{split}
\end{equation}
where $\overleftarrow{\prod_{j=1}^d}A_j=A_d...A_1$ is a product with decreasing indexes for non-commuting operators. 
More generally, for $\alpha=s,g$, we denote $ \tilde{U}_{\alpha}(t+h,t)=\overleftarrow{\prod_{j=1}^{d}}\hat{U}_{j,\alpha}(t+h,t)$ with $\hat{U}_{j,\alpha}(t+h,t)=\exp(-i \hat{C}_{j,\alpha}(t+h,t) \hat{D_j})$ and $\hat{C}_{j,s}(t+h,t)=h\hat{c}_j(t+h)$, $\hat{C}_{j,g}(t+h,t)=\int_t^{t+h}\hat{c}_j(s)ds$. Then, we define $\ket{\tilde{f}_\alpha}_t$ the qubit state given by the following recursive evolution:
\begin{equation}
\begin{split}
    \ket{\tilde{f}_{\alpha}}_{t+h}&=\tilde{U}_{\alpha}(t+h,t)\ket{\tilde{f}_{\alpha}}_{t}, \\
    \ket{\tilde{f}_{\alpha}}_{t=0}&=\ket{f_0},
\end{split}
\label{evolution product formula}
\end{equation}
where $t\in[0,T]$ and $h>0$. After a number $L$ of time-steps, one has prepared:
\begin{equation}
\label{eq:L time step evolution}
    \ket{\tilde{f}_{\alpha}}_T=\prod_{l=1}^L\hat{U}_{\alpha}(\frac{lT}{L},\frac{(l-1)T}{L})\ket{f_0}.
\end{equation}

The number $L$ of time-steps needed to reach a given error $\epsilon >0$ between $\ket{\tilde{f}_{\alpha}}_T$ and the solution $\ket{f}_T$ of the transport equation Eq.\ref{Transport equation} is determined in the error analysis section \ref{sec : error analysis}.
The product formula preserves the unitarity of the evolution which guarantees the $L^2$ stability of the quantum numerical scheme. This property is generally not achieved with other quantum numerical schemes based on an explicit Euler method, or using methods based on Dyson series expansion of the time-ordered evolution operator.

\section{Quantum numerical scheme}
\label{sec: quantum numerical scheme}

\begin{figure}
\centering
\begin{adjustbox}{width=1.2\textwidth}
\begin{quantikz}
\ket{0}\text{ }
&\gate[5,nwires={3,4},style={rounded
corners},disable auto height]{\begin{matrix}
\text{Quantum State Preparation:}\\ \text{}\\\text{Walsh Series Loader}\\\text{Fourier Series Loader}\\\text{Polynomial Series Loader}\\\text{...}\end{matrix}}  \slice{$\ket{f_0}$}&\gate[5,nwires={3,4},style={rounded
corners},disable auto
height]{\begin{matrix}
\text{Evolution:}\\\text{}\\\text{Diagonal unitaries and QFTs}\end{matrix}}  \slice{$\ket{f}_T$}&\gate[5,nwires={3,4},style={rounded
corners},disable auto
height]{\begin{matrix}
\text{Measurement protocol:}\\ \text{}\\\text{Hadamard test}\\\text{Swap test}\\\text{Quantum Amplitude Estimation}\\\text{...}\end{matrix}} &\\
 \ket{0}\text{ }& \qw & \qw & \qw & \\
 &&&&&&&&&&&\\
  & && && && &&&&& \\
\ket{0}\text{ } &\qw&\qw &\qw& & 
\end{quantikz}
\end{adjustbox}
\caption{Scheme of the quantum numerical scheme for the solution of a $d$-dimensional transport equation with position- and time-dependent coefficients $\hat{c}_j$.}
\label{fig:quantum numerical scheme}
\end{figure}
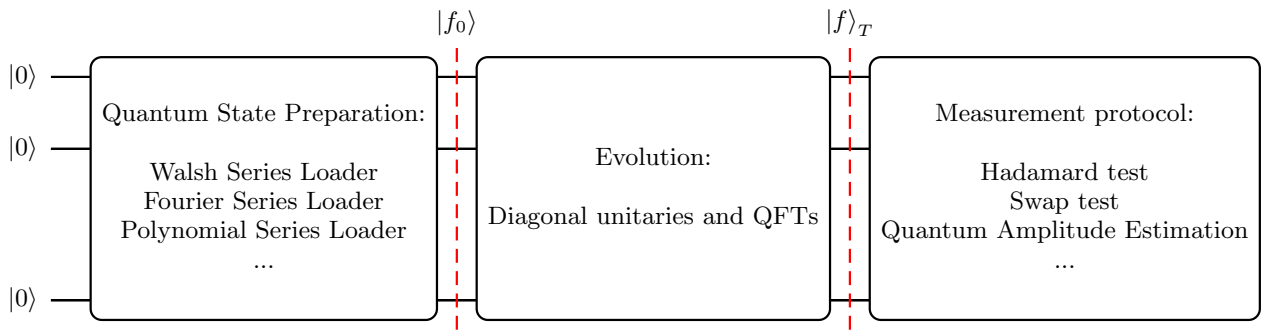

Solving transport equations on digital quantum computers requires three steps: 
\begin{itemize}
    \item Initialization : The initial condition is encoded into a qubit state:
    \begin{equation}
        \ket{0}^{\otimes n}\rightarrow \ket{f_0}.
    \end{equation}
    \item Evolution: The initial qubit state $\ket{f_0}$ evolves under a sequence of quantum gates that reproduce the desired evolution up to a time $T$: 
    \begin{equation}
        \ket{f_0}\rightarrow\ket{f}_T.
    \end{equation}
    \item Measurement of observables of interest: the prepared qubit state is used to measure a quantity of interest:
    \begin{equation}\ket{f}_T\rightarrow \leftindex_T{\bra{f}}\hat{O}\ket{f}_T.\end{equation}
\end{itemize}

In the following, we provide additional details on these three steps.

\subsection{Quantum state preparation}

The first step can be performed using a quantum state preparation algorithm. Exact preparations of arbitrary $n$-qubit states require quantum circuits made of $O(2^n)$ quantum gates, which is optimal \cite{sun2023asymptotically}. Without ancilla qubits\footnote{Ancilla qubits are additional qubits that help with the design of the quantum circuits.}, the optimal circuit depth\footnote{The depth of a quantum circuit refers to the number of layers of primitive quantum gates, whereas the circuit size counts the total number of such gates.} scales as $O(2^n/n)$ and, using $O(2^n)$ ancilla qubits, the depth can be reduced to $O(n)$ \cite{zhang2022quantum,yuan2023optimal}. To overcome these exponential scalings, it is possible to leverage the regularities of the initial function $f_0$. For instance, since $f_0$ is differentiable, one can represent $f_0$ in terms of Walsh series to obtain a precise approximation of $\ket{f_0}$ up to an approximation that depends on the maximum value of the first derivatives of $f_0$\cite{zylberman2024efficient,zylberman2025efficient}. Similarly, Fourier or Polynomial series can be leveraged to perform efficiently the initialization step when $f_0$ is regular enough \cite{moosa2023linear,gonzalez2024efficient,rosenkranz2025quantum}. Quantum circuits associated with the efficient preparation of qubit states depending on differentiable functions are presented in appendix \ref{Appendix: quantum state preparation}.    

\subsection{Evolution}

The evolution of the initial qubit state is performed through the unitary operators $\hat{U}_{j,\alpha}$ defined by the product formula Eq.(\ref{product formula}).
First, notice that the $j$-th derivative operator $\hat{D}_j$ is diagonalizable by the Quantum Fourier Transform (QFT) algorithm applied on the $j$-th register of qubits:
\begin{equation}
    \hat{D}_j= \widehat{QFT}_j^{-1}(\sum_{X_j}d_j(X_j)\ket{X_j}\bra{X_j}) \widehat{QFT}_j,
\end{equation}
where $\widehat{QFT}_j$ is the QFT acting on the $j$-th register and $d_j(X_j)=\frac{2}{\Delta x_j}\sum_{q=0}^p a_q\sin(2\pi q X_j)$ is one of the eigenvalues of the operator $\hat{D}_j$. Consequently, each operator $\hat{U}_{j,\alpha},$ is diagonalized by the QFT acting on the $j$-th register: 
\begin{equation}
\begin{split}
    \hat{\Lambda}_{j,\alpha}(t+h,t)&\equiv \widehat{QFT}_j\hat{U}_{j,\alpha}(t+h,t)\widehat{QFT}_j^{-1}\\&=\exp \big(-i \hat{C}_{j,\alpha}(t+h,t)d_j(\hat{X}_j)\big),
\end{split}
\label{eq: diag unitaries}
\end{equation}
where $\hat{X}_j=\sum_{X_j}X_j\ket{X_j}\bra{X_j}$ is the $j$-th position operator.

\begin{figure}
\centering
\begin{quantikz}
\ket{X_1}\text{ }
&\gate[1]{\widehat{QFT}_1}&\gate[1]{\hat{\Lambda}_{1,\alpha}}&\gate[1]{\widehat{QFT}_1^{-1}} &\qw \\
\ket{X_2}\text{ }
&\gate[1]{\widehat{QFT}_2}&\gate[1]{\hat{\Lambda}_{2,\alpha}}&\gate[1]{\widehat{QFT}_2^{-1}} &\qw \\
& & \vdots &&&&& \\
\ket{X_d}\text{ }
&\gate[1]{\widehat{QFT}_d}&\gate[1]{\hat{\Lambda}_{d,\alpha}}&\gate[1]{\widehat{QFT}_d^{-1}} &\qw
\end{quantikz}
\caption{Scheme of the quantum circuit associated with one time-step of the quantum numerical scheme for a $d$-dimensional transport equation with position-independent $\hat{c}_j$. In this case, the operators $\hat{\Lambda}_{j,\alpha}$ are diagonal operators acting on a single position register.}
\label{fig:schemeqc_position_independent}
\end{figure}
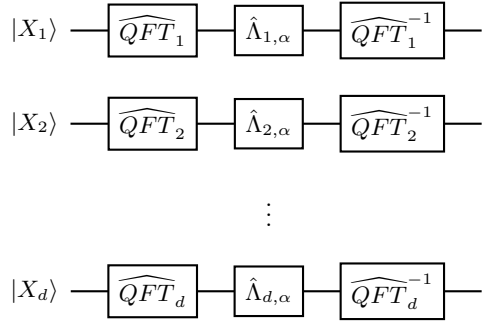

Therefore, for the $d$-dimensional transport equation, the evolution step is composed of $L\,d$ operators $\hat{U}_{j,\alpha}$, each of them being implementable with a diagonal unitary operator $\hat{\Lambda}_{j,\alpha}$ acting on the $n=n_1+...+n_d$ qubits, one QFT and one inverse QFT. Figure \ref{fig:schemeqc} presents a scheme of the quantum circuit for one time step for position-dependent $c_j$ coefficients. In the particular case where the $c_j$ coefficients are position-independent (and potentially time-dependent), each diagonal operator $\hat{\Lambda}_{j,\alpha}$ acts only on the $j$-th register of qubits, opening the possibility to parallelize the implementation of the QFTs and the diagonal operators (see Figure \ref{fig:schemeqc_position_independent}). The efficient quantum circuits for diagonal operators depending on differentiable functions are presented in appendix \ref{subsec: qc for diagonal unitaries}.

\begin{figure}
\centering
\begin{adjustbox}{width=\textwidth}
\begin{quantikz}
\ket{X_1} \text{ }
&\gate[1]{\widehat{QFT}_1}&\gate[4,nwires={3}]{\hat{\Lambda}_{1,\alpha}}&\gate[1]{\widehat{QFT}_1^{-1}} &\gate[4,nwires={3}]{\hat{\Lambda}_{2,\alpha}} &\qw & \qw & \hdots &&\qw &\gate[4,nwires={3}]{\hat{\Lambda}_{d,\alpha}} &\qw &\qw\\
 \ket{X_2}\text{ }& \qw &\qw & \gate[1]{\widehat{QFT}_2}& \qw &\gate[1]{\widehat{QFT}_2^{-1}} & \qw & \hdots && \qw & \qw & \qw &\qw  \\
 & && && &&\ddots&&&&& \\
 \ket{X_d} \text{ }&\qw&\qw &\qw& \qw & \qw &\qw &\hdots && \gate[1]{\widehat{QFT}_d} & \qw & \gate[1]{\widehat{QFT}_d^{-1}} &\qw
\end{quantikz}
\end{adjustbox}
\caption{Scheme of the quantum circuit associated with one time-step of the quantum numerical scheme for a $d$-dimensional transport equation with position-dependent velocity functions $\hat{c}_j$. The operators $\hat{\Lambda}_{j,\alpha}$ are diagonal operators acting on the $d$ registers.}
\label{fig:schemeqc}
\end{figure}
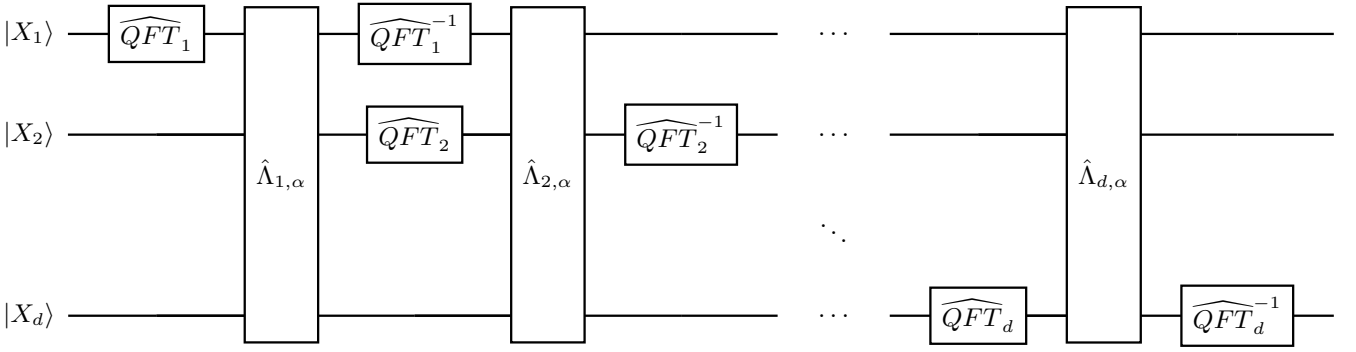

\subsection{Measurement of observables}

Sampling the $2^n$ components of the $n$-qubit state state $\ket{f}_T$ would require a prohibitively high number of repetitions of the numerical scheme --- typically one would need at least $O(2^n)$ measurements to reconstruct the amplitude of the state. Another approach consists in estimating the average value $ \leftindex_T{\bra{f}}\hat{O}\ket{f}_T$ of an observable $\hat{O}$ using an Hadamard test, a SWAP test, or a quantum amplitude estimation protocol \cite{brassard2000quantum,nielsen2010quantum,shang2024estimating,grinko2021iterative,rall2023amplitude,giurgica2022low}. For the Hadamard and SWAP tests, one needs to repeat the preparation of $\ket{f}_T$ a number $O(1/\epsilon^2)$ of times, where $\epsilon>0$ is the error associated with the measurement protocol, while for the amplitude estimation protocol the scaling is quadratically smaller as $O(1/\epsilon)$\footnote{In general, these measurement protocols produce a probabilistic outcome.} These protocols and the associated quantum circuits are presented in appendix \ref{subsec: qc for measurement}.

In the following, we provide examples of observables $\hat{O}$ that correspond to quantities of interest.

For instance, one can estimate the solution $f$ at time $T$ at a given discrete position $\vec{X}$ with the following observable:
\begin{equation}        \hat{O}_{\vec{X}}=\ket{\vec{X}}\bra{\vec{X}}\longrightarrow  \leftindex_T {\bra{f}}\hat{O}_{\vec{X}}\ket{f}_T=\frac{|f(\vec{X},T)|^2}{\|f_0\|_{2,N}^2},
\end{equation}
where the normalization constant $\|f_0\|_{2,N}$ can be evaluated as
$\|f_0\|_{2,N}\simeq\sqrt{N}\|f_0\|_{L^2}=\sqrt{N}\sqrt{\int_{[0,1]^d}|f_0(\vec{x})|^2dV}$ for $N$ large enough. For small $N$, the normalization factor can be directly computed using a classical computer. Other integrals involving $f$ and moments of the position variables $\braket{x_1^{k_1}...x_d^{k_d}}$ of any order $\vec{k}=(k_1,...,k_d)\in \mathbb{N}^d$ can be estimated with the following observables:
\begin{equation}
\begin{split}
    \hat{O}_s&=\ket{s}\bra{s}\longrightarrow \leftindex_T {\bra{f}}\hat{O}_s\ket{f}_T= |\sum_{\vec{X}}f(\vec{X},T)|^2/(N\|f_0\|_{2,N}^2) \overset{N\rightarrow +\infty}{=} \frac{\big ( \int_{[0,1]^d}f(X,T)dV \big )^2}{\|f_0\|_{L^2}^2},
     \\
    \hat{O}_{\vec{k}}&=\sum_{\vec{X}}X_1^{k_1}...X_d^{k_d}\ket{\vec{X}}\bra{\vec{X}}\longrightarrow \leftindex_T {\bra{f}}\hat{O}_{\vec{k}}\ket{f}_T=\sum_{\vec{X}}X_1^{k_1}...X_d^{k_d}\frac{|f(\vec{X},T)|^2}{\|f_0\|_{2,N}^2}\overset{N\rightarrow +\infty}{=}  \frac{\int_{[0,1]^d}x_1^{k_1}...x_d^{k_d}|f(\vec{x},T)|^2dV }{\|f_0\|_{L^2}^2},
    \\
    \hat{O}_{\vec{k},\text{bis}}&=\ket{f}_T\big (\frac{1}{\mathcal{N}_{\vec{k}}}\sum_{\vec{X}}X_1^{k_1}...X_d^{k_d}\bra{\vec{X}}\big )\longrightarrow \leftindex_T {\bra{f}}\hat{O}_{\vec{k},\text{bis}}\ket{f}_T=\sum_{\vec{X}}X_1^{k_1}...X_d^{k_d}\frac{f(\vec{X},T)}{\mathcal{N}_{\vec{k}}\|f_0\|_{2,N}}\overset{N\rightarrow +\infty}{=}  C_{\vec{k}}\frac{\int_{[0,1]^d}x_1^{k_1}...x_d^{k_d}f(\vec{x},T)dV }{\|f_0\|_{L^2}}, 
\end{split}
\label{eq: observables}
\end{equation}

where $\ket{s}=(1/\sqrt{N})\sum_{\ket{\vec{X}}}\ket{\vec{X}}$ is the full superposition state, $\mathcal{N}_{\vec{k}}=\sqrt{\sum_{\vec{X}}\big(X_1^{k_1}\hdots X_d^{k_d}\big)^2}$ is a normalization constant, $C_{\vec{k}}=1/\sqrt{(2k_1+1)\hdots(2k_d+1)}$ is a constant coming from the term $\mathcal{N}_{\vec{k}}$ and $\|f_0\|_{L^2}=\sqrt{\int_{[0,1]^d}|f_0(x)|^2dV}$. The limit $N \rightarrow +\infty$ means that $\forall j, N_j\rightarrow +\infty$. Notice that these observables are either non-unitary diagonal operators, or they can be prepared through quantum state preparation for differentiable functions as presented in references \cite{zylberman2025efficient,zylberman2024efficient,moosa2023linear}.
As shown in the application section \ref{sec: application}, moments of the form $\braket{x_1^{k_1}...x_d^{k_d}}$ are particularly interesting for the collisionless Boltzmann equation, since they link the kinetic distribution function of the particles to fluid quantities, such as the fluid density, the fluid velocity or the energy of the particles. The average position $\braket{x_j}$ is also particularly relevant when solving non-linear Hamiltonian ordinary differential equations via their Liouville equation. As shown with the Lotka-Volterra example in section \ref{sec: non-linear ODE solver}, the average position over one axis directly corresponds to an estimation of the solution of the non-linear ODE system. Moreover, when the solution is localized enough, a single measurement of the $n$ qubits encoding $\ket{f}_T$ suffices to estimate the average position.

\section{Error and complexity analysis}
\label{sec : error analysis}
The quantum numerical scheme deals with seven different sources of error\footnote{
Hardware noise is dependent on the underlying technology of the quantum processor unit and is not discussed here.}:
\begin{itemize}
    \item Quantum circuit approximation of the quantum state preparation 
    \item Space discretization 
    \item Product formula approximation
    \item Quantum circuit approximation of the evolution (diagonal unitary approximation)
    \item Observable discretization (quadrature)
    \item Quantum circuit approximation of the observable
    \item Statistical measurement error
\end{itemize}

We denote by $O_T$ the scalar quantity one wants to estimate and $\langle \tilde{O}_T\rangle$ the quantity obtained after the measurement protocol from the implemented operator $\tilde{\hat{O}}$ and the implemented quantum state that we define as:
\begin{equation}
    \ket{\tilde{\tilde{\tilde{f}}}_\alpha}_T=\prod_{l=1}^L\tilde{\hat{U}}_{\alpha}(\frac{lT}{L},\frac{(l-1)T}{L})\ket{\tilde{f}_0},
\end{equation}
where $\tilde{\hat{U}}_{\alpha}$ is a quantum circuit implementation of the operator $\hat{U}_{\alpha}$ for which an optional quantum circuit approximation has been performed. Similarly, $\ket{\tilde{f}_0}$ is the encoding of the initial condition $f_0$ for which an optional quantum circuit approximation has been performed (the exact encoding of $f_0$ is $\ket{f_0}$). Additionally, we note $\ket{\tilde{\tilde{f}}_\alpha}_T=\prod_{l=1}^L\tilde{\hat{U}}_{\alpha}(\frac{lT}{L},\frac{(l-1)T}{L})\ket{f_0}$ and we notice that the difference in two-norm of $\ket{\tilde{\tilde{f}}_\alpha}_T$ and $\ket{\tilde{\tilde{\tilde{f}}}_\alpha}_T$ is $\|\ket{\tilde{f}_0}-\ket{f_0}\|_{2,N}$. By using the triangular inequality and the Cauchy-Schwartz inequality, the difference $|O_T-\langle\tilde{O}_T\rangle|$ can be bounded by the seven sources of error as:   
\begin{equation}
\begin{split}
    |O_T- \langle\tilde{O}_T\rangle|&\leq \overbrace{|O_T-\leftindex_T{\bra{f}}\hat{O}\ket{f}_T|}^{\text{Quadrature error}}+\overbrace{|\langle\tilde{O}_T\rangle-\leftindex_T{\bra{\tilde{\tilde{\tilde{f}}}_\alpha}}\tilde{\hat{O}}\ket{\tilde{\tilde{\tilde{f}}}_\alpha}_T|}^{\text{Statistical measurement error}}+\overbrace{\|\hat{O}-\tilde{\hat{O}}\|_2}^{\text{q.c. approx.}}\\& +2\|\hat{O}\|_{2}\big(\underbrace{\|\ket{\tilde{\tilde{f}}_\alpha}_T-\ket{\tilde{f}_\alpha}_T\|_{2,N}}_{\text{q.c. approximation}}+\underbrace{\|\ket{\tilde{f}_{\alpha}}_T-\ket{\tilde{f}}_T\|_{2,N}}_{\text{Product formula error}}+\underbrace{\|\ket{\tilde{f}}_T-\ket{f}_T\|_{2,N}}_{\text{Space discretization}}+\underbrace{\|\ket{\tilde{f}_0}-\ket{f_0}\|_{2,N}}_{\text{Q.S.P error}}\big),
\label{Eq:error accumulation}
\end{split}
\end{equation}
where we remind the reader that $\|.\|_{2,N}$ is the vector $2$-norm of a vector of $N$ components, $\ket{f}_T$ is the exact $n$-qubit encoding of the solution at time $T$, $\ket{\tilde{f}}_T$ is the $n$-qubit encoding of the solution of the ODE system Eq.(\ref{ODE}) obtained after space discretization and $\ket{\tilde{f}_\alpha}_T=\prod_{l=1}^L\hat{U}_{\alpha}(\frac{lT}{L},\frac{(l-1)T}{L})\ket{f_0}$ is the $n$-qubit state obtained after the $\alpha=g,s$ product formula approximation. The notation ``q.c. approx." stands for quantum circuit approximation and ``Q.S.P." for quantum state preparation. Notice that some errors vanish in some particular cases. For instance, if the observable is not a discretized (on the spacial grid) quantity, the quadrature error vanishes; or if some quantum circuits are implemented without approximations.

The quantum circuit approximations for diagonal operators and quantum state preparation are discussed in appendix \ref{appendix:qc}. The quadrature error depends highly on the observable $\hat{O}$ and requires basic algebra for Riemann quadrature that converges linearly with the number of points $N$, i.e., exponentially with the number of qubits. In the following, we bound the space discretization error and we provide novel numerical analysis to bound the product formula approximation with vector norm scalings.

\subsection{Space-discretization error}
\label{subsec: space discretization error}

Space discretization induces an error between the solution $\ket{f}_t$  of the transport equation Eq.(\ref{Transport equation}) and the solution $\ket{\tilde{f}}_t$ of the system of ODEs, Eq.(\ref{ODE}). This error depends directly on the order of the central finite difference scheme. First, we characterize the action of a discrete operator $\hat{D}_j$ acting on a vector depending on a continuously differentiable function:

\begin{lemma}
 Let $n\in\mathbb{N}^*$ and $g$ be $(2p+1)$-differentiable function defined on $[0,1]^d$. The discrete derivative operator $\hat{D}_j$ of order $2p$, defined by Eq.(\refeq{2p discrete derivative operator}) and applied on the normalized $n$-qubit state $\ket{g}=\frac{1}{\mathcal{N}}\sum_{\vec{X}}g(\vec{x})\ket{\vec{X}}$, approximates the first order derivative of $g$ as:
\begin{equation}
    i\hat{D}_j\ket{g}=\ket{\partial_{x_j}g}+O((\Delta x_j)^{2p}).
\end{equation}
In particular, for the vector $2$-norm $||.||_{2,N}$:
\begin{equation}
    ||\hat{D}_j\ket{g}||_{2,N}\leq ||\ket{\partial_{x_j}g}||_{2,N}+\tilde{C}_p(\Delta x_j)^{2p},
\end{equation}
where  $\ket{\partial_{x_j}g}=\frac{1}{\mathcal{N}}\sum_{\vec{X}}\partial_{x_j}g(\vec{X})\ket{\vec{X}}$ and $\tilde{C}_p$ is a constant independent of $n$, depending only on $p$, the $L^2$ norm of $g$ and the $L^2$ norm of the $(2p+1)-j$-th derivative of $g$.
\label{lemma discrete derivative operator}
\end{lemma}

\begin{proof}
The proof is a direct consequence of Taylor's theorem:
For any $p$-continuously differentiable function $g$ scale-valued or matrix valued defined on an interval $[a,b]$, one has:
\begin{equation}
    g(b)=\sum_{j=0}^{p-1}\frac{g^{(j)}(a)}{j!}(b-a)^j+\int_a^b\frac{g^{(k)}(s)}{(p-1)!}(b-s)^{p-1}ds.
\label{Taylor formula}
\end{equation}
More details on the constant $\tilde{C}_p$ are given in appendix  \ref{appendix: proof of lemma finite difference}.
\end{proof}

Now it is possible to fully characterize the space discretization error. The following theorem states that the error coming from an equidistant discretization of each axis and a $2p$ accurate central finite difference scales linearly with time $t$, the maximum value of the $c_j$ functions and $(\Delta x_j)^{2p}$:

\begin{theorem}
\label{thm_discretization_error}
The vector norm error between the real-space encoding of the solution $\ket{f}_t$ of the $d$-dimensional transport equation Eq.(\ref{Transport equation}) and the vector $\ket{\tilde{f}}_t$ solution of ODE system Eq.(\ref{ODE}) using a finite difference scheme of order $2p$ at time $t\in[0,T]$ is bounded as:
\begin{equation}
    ||\ket{f}_t-\ket{\tilde{f}}_t||_{2,N}\leq t K \sum_{j=1}^d ||c_j||_\infty (\Delta x_j)^{2p},
\end{equation}
where $\ket{f}_t=\frac{1}{||f_0||_{2,N}}\sum_{\vec{X}}f(\vec{X},t)\ket{\vec{X}}$ and, $K$ is a constant depending on $p$ and the maximum value over $j\in\{0,...,d\}$ and time of the $(2p+1)$ derivative of $f_0$ with respect to axis $j$.
\end{theorem}

The proof, based on Lemma \ref{lemma discrete derivative operator} and the variation of parameter formula, is given in appendix \ref{appendix: proof of thm discretization error}.

The space discretization error scales with $(\Delta x_j)^{2p}$, where $\Delta x_j=1/2^{n_j}$ for a uniform discretization. Consequently, to reach an error $\epsilon>0$ at the final time $T$, one needs to take a number $n_j$ of qubits associated with the real space encoding of the $j$-th axis as
\begin{equation}
\label{qubit scaling}
n_j=\frac{\log_2(TdK||c_j||_\infty/\epsilon)}{2p}.
\end{equation}

\begin{figure}[h]
\centering
\includegraphics[scale=0.6]{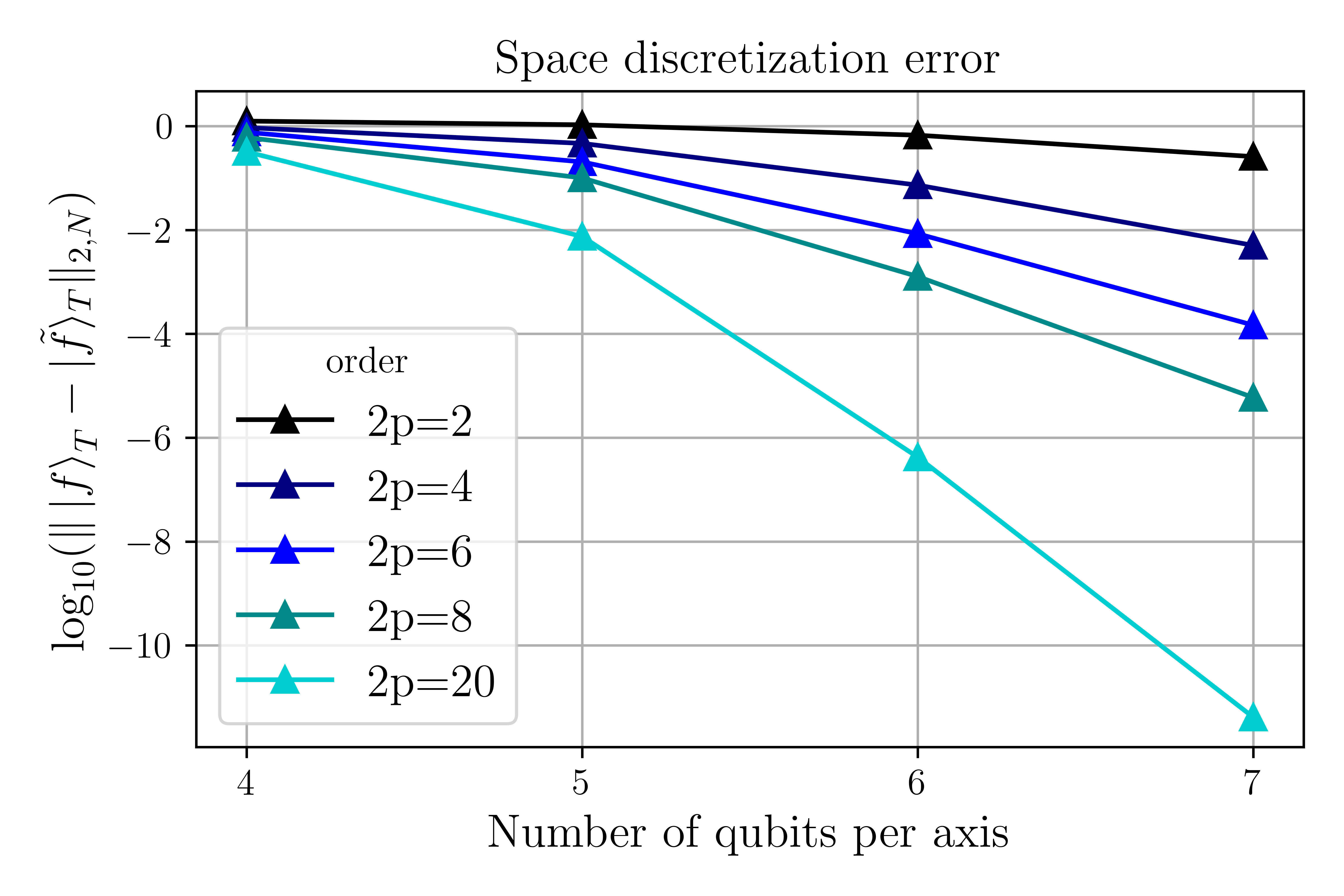}
\caption{Discretization error $||\ket{f}_T-\ket{\tilde{f}}_T||_{2,N}$ as a function of the number of qubits per axis for the transport equation of dimension $d=2$ defined in section \ref{subsec: space discretization error} with $c_1(x_2)=(x_2-0.5), c_2(x_1)=(qE_0/m)(x_1-0.5)$ and $qE_0/m=-1$. The simulation time is $T=2\pi$ and the initial condition is $f_0(x_1,x_2)\propto e^{-((x_1-\mu_1)^2+(x_2-\mu_2)^2)/(2\sigma^2)}$ with $\mu_1=-0.1,\mu_2=0.1$ and $\sigma=1/(10\sqrt{2})$.}
\label{fig: discretization error}
\end{figure}

In practice, one can estimate numerically the value of $K$ by using Equation \ref{inequality_on_discretization} or by performing a linear regression on the discretization error for a small number of qubits, i.e., for cases that can be computed classically.  In Figure \ref{fig: discretization error}, the vector norm error is displayed for a two dimensional transport equation defined in section \ref{subsec: Collisionless Boltzmann equation} and for different order of the central finite differences, confirming that the discretization error decreases exponentially with the number of qubits per axis.

\subsection{Product formula approximation}
\label{subsec:Product formula approximation}

The study of product formula approximations has attracted a lot of attention in the past decade \cite{An2021,Childs2021,su2021nearly,layden2022first}. A direct expansion of the Trotter formula gives $\|\big(e^{i(\hat{H}_1+\hat{H}_2)\Delta t}-e^{i\hat{H}_1 \Delta t}e^{i\hat{H}_2 \Delta t}\big)\ket{\psi}\|_{2,N}= \frac{\Delta t^2}{2}\|[\hat{H}_2,\hat{H}_1]\ket{\psi}\|_{2,N}+O(\Delta t^3)$. In the literature, most of the scaling uses the operator norm inequality by bounding the commutator as $\|[\hat{H}_2,\hat{H}_1]\ket{\psi}\|_{2,N}\leq \|[\hat{H}_2,\hat{H}_1]\|_{2}\leq 2\|\hat{H}_1\|_{2}\|\hat{H}_2\|_{2}$, where $\|.\|_2$ is the spectral norm \cite{Childs2021}. However, in the case of the transport equation, the summands $\hat{H}_j$ depend on the discrete derivative operator such as $\|\hat{H}_j\|_{2}=\|c_j\hat{D}_j\|_2=O(1/\Delta x_j)=O(2^{n_j})$, resulting in exponential-with-$n$ error bounds\footnote{Note that only the norm of $\hat{D}_j$ is asymptotically unbounded since the functions $c_j$ are assumed to be bounded: $||\hat{c}_j(t)||_2 \leq ||c_j||_{\infty}=\sup_{\vec{x}\in [0,1[^d,t\in[0,T]}|c_j(\vec{x},t)|=O(1)$.}. The following theorem states the operator norm inequality, expressed in the notation of this paper for the standard and generalized product formula defined Eq.\ref{product formula}:

\begin{theorem}[Operator norm inequality]
\label{thm_product_formula_error_operator_norm}
    The error between the solution $\ket{\tilde{f}}_t$ of the ODE system Eq.(\ref{ODE}) and the vector $\ket{\tilde{f}_{\alpha}}_t$, with $\alpha=s,g$, solution of Eq.(\ref{evolution product formula})  using a finite difference scheme of order $2p$, with $p\geq 2$, at the final time $T$ and after $L>0$ time-steps is bounded as:
\begin{equation}
\begin{split}
    ||\ket{\tilde{f}_{g}}_T-\ket{\tilde{f}}_T||_{2,N}& \leq \alpha_g\ \frac{T^2}{L},
\\
    ||\ket{\tilde{f}_{s}}_T-\ket{\tilde{f}}_T||_{2,N}&\leq \alpha_{s} \frac{T^2}{L} +\beta_{s}\frac{T^3}{L^2}, 
\end{split}
\label{Eq: Scaling}
\end{equation}
where
\begin{equation}
\begin{split}
    \alpha_g&=\sum_{j=1}^{d-1}\sum_{m=j+1}^d \|c_j\|_{\infty}  \|c_m\|_{\infty} \|\hat{D}_j\|_2\|\hat{D}_m\|_2 , \\
    \alpha_{s}&=\alpha_g+\frac{1}{2}\sum_{j=1}^d\|\partial_tc_j\|_{\infty} \|\hat{D}_j\|_2 ,\\
    \beta_{s}&=\frac{1}{3} \sum_{j=1}^{d-1}\sum_{m=j+1}^d \|c_j\|_{\infty}\|\partial_t c_m\|_{\infty} \|\hat{D}_j\|_2\|\hat{D}_m\|_2 .
\end{split}
\end{equation}

\end{theorem}

The proof is given in appendix \ref{appendix: proof of trotterization error}.

Notice that the definition of the discrete derivative operator implies that $\|\hat{D}_j\|_2\leq2C_p/\Delta x_j$, where $C_p=\sum_{q=0}^pa_q$. Therefore, the constants $\alpha_g,\alpha_s$ and $\beta_s$ scale as $O(\max_j\frac{1}{\Delta x_j^2})=O(\max_j4^{n_j})$. Using these operator norm inequalities, the number of time-steps required to achieve an error $\epsilon>0$ should scale as:
\begin{equation}
    L=O(\frac{T^2}{\epsilon}\max_j4^{n_j}).
\label{eq:time_step operator norm}
\end{equation}
In the following, we generalize the computations of \cite{An2021} to transport equations\footnote{Reference \cite{An2021} is the first to  introduce vector norm analysis for quantum algorithm. They address the problem of solving the Schrödinger equation with a bounded potential, i.e., in the case where $\hat{H}_1$ is asymptotically unbounded and $\hat{H}_2$ is bounded, an assumption used in the proofs of their results. Our results generalize their approach since all the summands in the Hermitian operator $\sum_{j=1}^d \hat{c}_j(t)\hat{D}_j$ are asymptotically unbounded.}, and overcome these exponential-with-$n_j$ scalings by considering the action of the operator $\hat{D}_j$ on the qubit state encoding the solution $\ket{f}_t$ (see Lemma (\ref{lemma discrete derivative operator})), deriving a novel bound in vector norm with terms of the form $\|[\hat{c}_j\hat{D}_j,\hat{c}_{j'}\hat{D}_{j'}]\ket{f}_t\|_{2,N}=O(1)$. These exponential-with-$n_j$ improvements are summarized in the following theorem:

\begin{theorem}[Vector norm inequality]
\label{thm_product_formula_error}
    The vector norm error between the solution $\ket{\tilde{f}}_t$ of the ODE system Eq.(\ref{ODE}) and the vector $\ket{\tilde{f}_{\alpha}}_t$, with $\alpha=s,g$, solution of Eq.(\ref{evolution product formula})  using a finite difference scheme of order $2p$, with $p\geq 2$, at the final time $T$ and after $L>0$ time-step is bounded as:

\begin{equation}
\begin{split}
    ||\ket{\tilde{f}_{g}}_T-\ket{\tilde{f}}_T||_{2,N}&\leq  \alpha_g' \frac{T^2}{L}+ r_g ,\\
    ||\ket{\tilde{f}_{s}}_T-\ket{\tilde{f}}_T||_{2,N}&\leq \alpha_s' \frac{T^2}{L}+ r_s,
\end{split}
\label{Eq: Scaling_bis}
\end{equation}
where $r_g,r_s$ contain higher order terms that are asymptotically negligible and

\begin{equation*}
\begin{split}    
\alpha_g'&=\frac{1}{2}  \sum_{j=1}^{d-1} \sum_{m=j+1}^d \bigg( \|c_j\|_{\infty} \|\partial_{x_j} c_m\|_{\infty} \max_{t\in[0,T]}\frac{ \|\partial_{x_m}f_t\|_{2,N}}{\|f_0\|_{2,N}}
     +\|c_m\|_{\infty} \|\partial_{x_m} c_j\|_{\infty} \max_{t\in[0,T]}\frac{ \|\partial_{x_j}f_t\|_{2,N}}{\|f_0\|_{2,N}} \bigg) \\ & \leq\frac{d(d-1)}{2}\max_{1\leq m<j\leq d, t\in[0,T]} ||c_j||_{\infty}||\partial_{x_j}c_m||_{\infty}\frac{||\partial_{x_m}f_t||_{2,N}}{||f_0||_{2,N}}, \\
\alpha_s'&=\alpha_g' +\frac{1}{2}\sum_{j=1}^d\|\partial_tc_j\|_{\infty} \max_{t\in[0,T]}\frac{\|\partial_{x_j}f_t\|_{2,N}}{\|f_0\|_{2,N}} \\ & \leq \alpha_g'+ \frac{d}{2}\max_{1\leq j \leq d, t\in [0,T]} ||\partial_tc_j||_{\infty} \frac{||\partial_{x_j} f_t||_{2,N}}{||f_0||_{2,N}},
\end{split}
\end{equation*}

where $\|\partial_{x_m}f_t\|_{2,N}=\sqrt{\sum_{\vec{X}}|\partial_{x_m}f(\vec{X},t)|^2}$ and $\|f_0\|_{2,N}=\sqrt{\sum_{\vec{X}}|f_0(\vec{X})|^2}$.
\end{theorem}

The proof of this theorem and the expressions for $r_g$ and $r_s$ are given in appendix \ref{appendix: proof of trotterization error}. 

Note that $||\partial_{x_m}f_t||_{2,N}/||f_0||_{2,N}$ converges toward  $||\partial_{x_m}f_t||_{L^2}/||f_0||_{L^2}=||\partial_{x_m}f_0||_{L^2}/||f_0||_{L^2}$ in the large $N$ limit, which depends only on $f_0$\footnote{The $L^2$ norm of any derivative of $f$ is constant over time, equal to its initial value at time $t=0$. See appendix \ref{appendix: proofs} for more details.}. Asymptotically, the coefficients $\alpha_g'$ and $\alpha_s'$ depend only on the dimension $d$ and different norms of the functions $c_j$ and $f_0$. They do not scale with the norms of the operators $\hat{D}_j$, i.e., with the factors $1/(\Delta x_j)$, improving the operator-norm inequality by a factor of $\max_j1/(\Delta x_j)^2=O(\max_j4^{n_j})$.

As a consequence, the number $L$ of time steps needed to reach a given error $\epsilon>0$ at time $T$ can be reduced by a factor $O(\max_j4^{n_j})$ compared to the scaling predicted by the operator norm inequality Eq.~\ref{eq:time_step operator norm}, by choosing $L$ as: 

\begin{equation}
    L=O(\alpha_{\mu}' \frac{T^2}{\epsilon}),
\end{equation}
with $\mu=s,g$. This scaling for $L$ implies that the terms $r_g$ and $r_s$ present in the vector norm inequalities are asymptotically negligible for $p\ge2$: $r_g=O(\epsilon^{3(1-1/2p)-1}/T^{1-1/2p})=O(\epsilon^{5/4})$ and $r_s=O(\epsilon^{3(1-1/2p)-1}/T^{1-1/2p})=O(\epsilon^{5/4})$ for $p\ge 2$. 

\begin{figure}[ht]
\centering
    \begin{subfigure}{0.47\textwidth}
        \includegraphics[width=\textwidth]{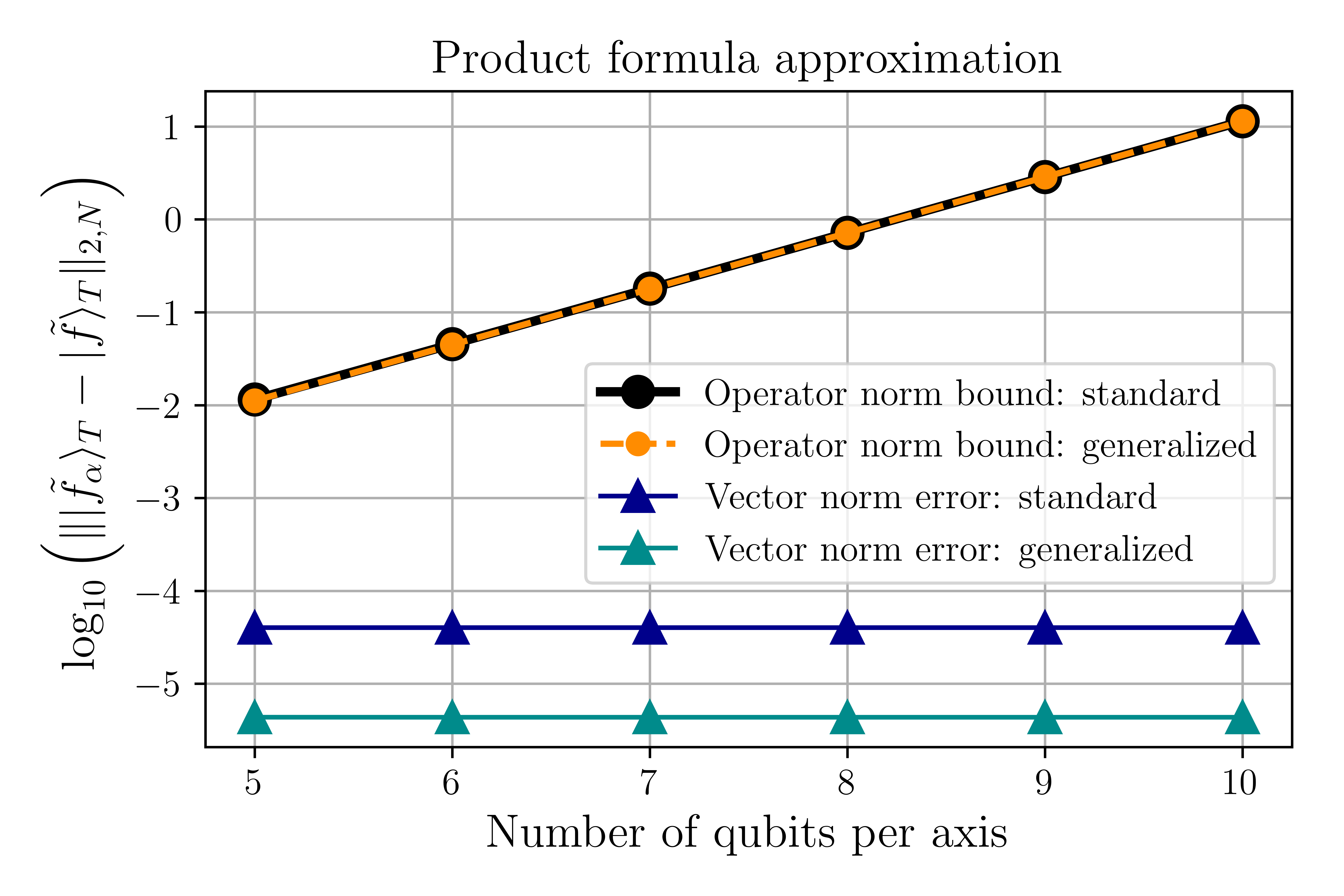}
        \caption{}
        \label{fig:trotter_error_with_n}
    \end{subfigure}
    \begin{subfigure}{0.47\textwidth}
        \includegraphics[width=\textwidth]{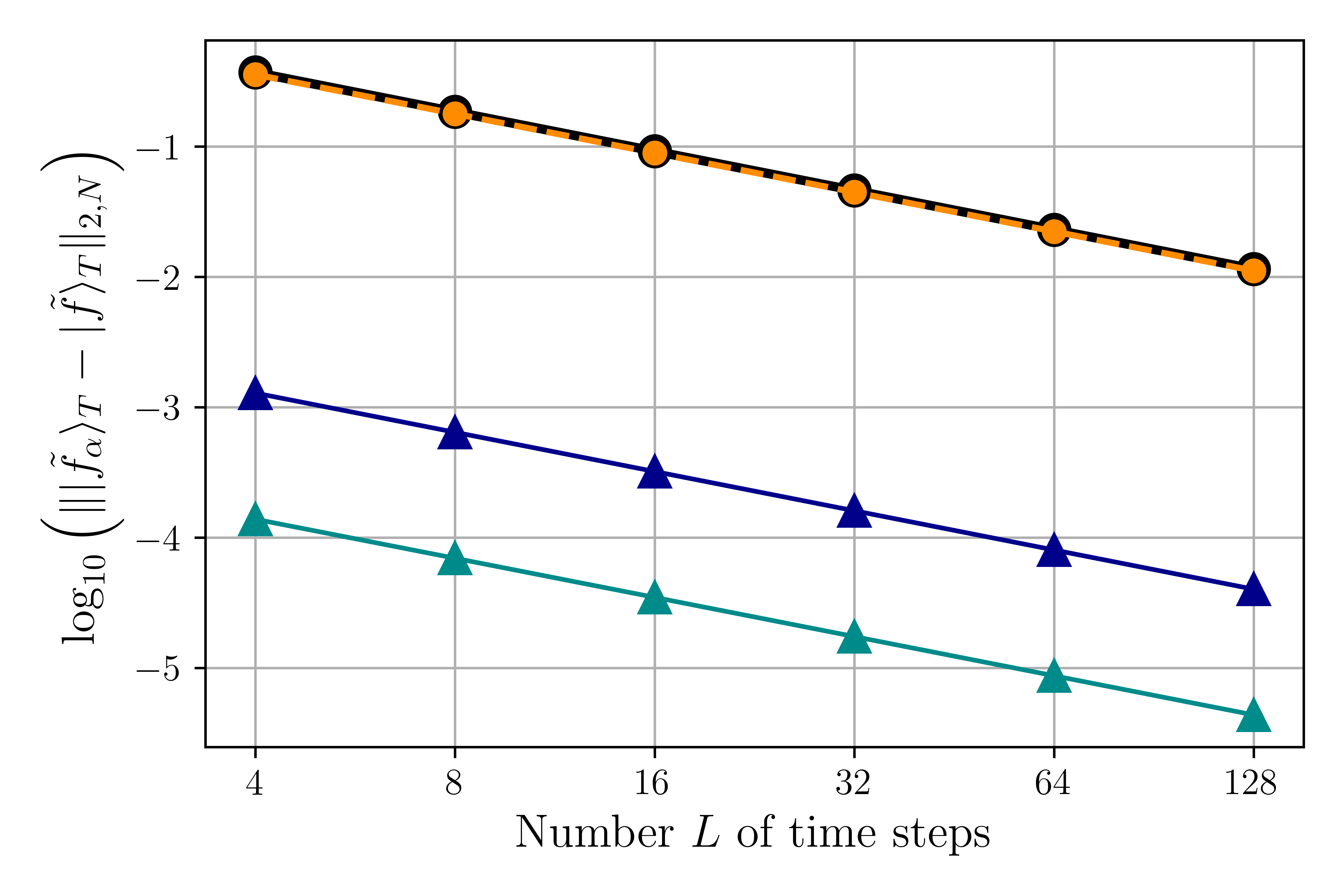}
        \caption{}
        \label{fig:trotter_error_with_L}
    \end{subfigure}
\caption{Product formula error $||\ket{\tilde{f}_\alpha}_T-\ket{\tilde{f}}_T||_{2,N}$ for the generalized product formula $\alpha=g$ and the standard product formula $\alpha=s$ as a function of the number of qubits per axis with $L=128$ (left) and, as a function of the number of time steps $L$ with $n=10$ (right). The simulations are performed for the transport equation of dimension $d=2$ defined in section \ref{subsec: Collisionless Boltzmann equation} with $c_1(x_2)=(x_2-0.5), c_2(x_1,t)=(qE_0/m)(x_1-0.5-ct)$ and parameters $qE_0/m=-1$, $c=1.6$. The numerical solution $|\ket{\tilde{f}_\alpha}_T$ is compared to the analytical solution of the transport equation computed via the method of characteristics. Final time is $T=0.025$ and the initial condition is $f_0(x_1,x_2)\propto e^{-((x_1-\mu_1)^2+(x_2-\mu_2)^2)/(2\sigma^2)}$ with $\mu_1=-0.1,\mu_2=0.1$ and $\sigma=1/(10\sqrt{2})$.}
\label{fig: product formula approximation}
\end{figure}

Scalings of the product formula approximations are illustrated in Figure \ref{fig:trotter_error_with_L} by solving a two dimensional transport equation defined in section \ref{subsec: Collisionless Boltzmann equation}. The vector norm of the difference between $\ket{\tilde{f}_\alpha}_T$ and $\ket{\tilde{f}}_T$ does not scale with the number of qubits as expected by the vector norm inequalities \ref{thm_product_formula_error} while the operator norm bounds scale exponentially with the number of qubits.

\subsection{Complexity analysis}
\label{subsec: complexity analysis}
The complexity analysis of the quantum numerical scheme can be divided into the quantum state preparation part, the evolution part and the number of times one needs to perform these two steps to perform a measurement protocol. 

The evolution has $L=O(d^2T^2/\epsilon)$ time-steps and each time step requires the implementation of $2d$ QFTs and $d$ diagonal unitaries. The amplitude estimation protocols \ref{fig: quantum amplitude estimation} need $O(1/\epsilon)$ repetitions to estimate the average value of an observable up to an accuracy $\epsilon>0$. Therefore, one has to implement $O(1/\epsilon)$ quantum state preparations, $O(d^3T^2/\epsilon^2)$ diagonal unitaries, $O(d^3T^2/\epsilon^2)$ QFTs and $O(1/\epsilon)$ observable operators $\hat{O}$. When the quantum circuits are approximated to reduce the circuit complexity, the error accumulates according to Equation \ref{Eq:error accumulation}. 
Let $C_{\text{QSP}},C_{\text{diag}},C_{\text{qft}},C_{\text{obs}}$ be the computational costs of the initialization, the most expensive diagonal unitaries, the quantum Fourier transform, and the implementation of the observable $\hat{O}$; then, the circuit size and depth\footnote{The quantum amplitude estimation protocols (see Fig.~\ref{fig: quantum amplitude estimation}) require at least one ancilla qubits to control these operations, which can be achieved with the same size and depth scaling.} scale as
\begin{equation}
   O\big (\frac{C_{\text{QSP}}}{\epsilon}+\frac{d^3T^2(C_{\text{diag}}+2C_{\text{QFT}})}{\epsilon^2}+ \frac{C_{\text{obs}}}{\epsilon} \big).
\label{Eq: complexity scaling}
\end{equation}

First, notice that the complexity scaling of the evolution step has been reduced by a factor $4^{n_j}$ when compared to the complexity based on the operator norm inequality (which guarantees an $\epsilon$ error of the product formula with
$L = O(d^2 T^2(\max_j 4^{n_j} )/\epsilon)$ time steps).

Then, the computational costs $C_{\text{qsp}}$ and $C_{\text{diag}}$ are bounded by the complexities of the exact implementations, as $C_{\text{qsp,exact}}=O(2^n)=O(T^{\frac{d}{2p}}d^{\frac{d}{2p}}/\epsilon^{\frac{d}{2p}})$,  $C_{\text{diag,exact}}=O(2^n)=O(T^{\frac{d}{2p}}d^{\frac{d}{2p}}/\epsilon^{\frac{d}{2p}})$. For the observable $\hat{O}$, exact unitary synthesis requires $O(4^n)$ gates but for the observable presented in Equation \ref{eq: observables}, diagonal operations and quantum state preparation suffice to implement them.

It is possible to improve these exponential-with-$n$ scalings by leveraging the fact that the diagonal unitaries and the quantum state preparation depend on bounded differentiable functions \footnote{Note that $\Delta t/\Delta x =O(\epsilon^{1-1/(2p)})$.}. As shown in references \cite{welch2014efficient,zylberman2024efficient,zylberman2025efficient} and in appendix \ref{appendix:qc}, they can be $\delta$-approximated with a circuit depth independent of the number of qubits $n$ using $M$-Walsh series or sparse Walsh series. The size of the quantum state preparation is linear with the number $n$ of qubits (see the Quantum State Preparation Theorem 8.1 of reference \cite{zylberman2025efficient}) and the size of the diagonal operations is independent of $n$ (see Theorem 4.1 and Table 1 of reference \cite{zylberman2025efficient}). However, the errors due to these quantum circuit approximations accumulate as the sum of the error of each individual diagonal operator:
\begin{equation}    \|\ket{\tilde{\tilde{f}}_\alpha}_T-\ket{\tilde{f}_\alpha}_T\|_{2,N}\leq\sum_{l=1}^{L}\sum_{j=1}^d\|\tilde{\hat{U}}_{j,\alpha}-\hat{U}_{j,\alpha}\|_2.
\end{equation}

\begin{figure}[ht]
\centering
\includegraphics[width=0.5\textwidth]{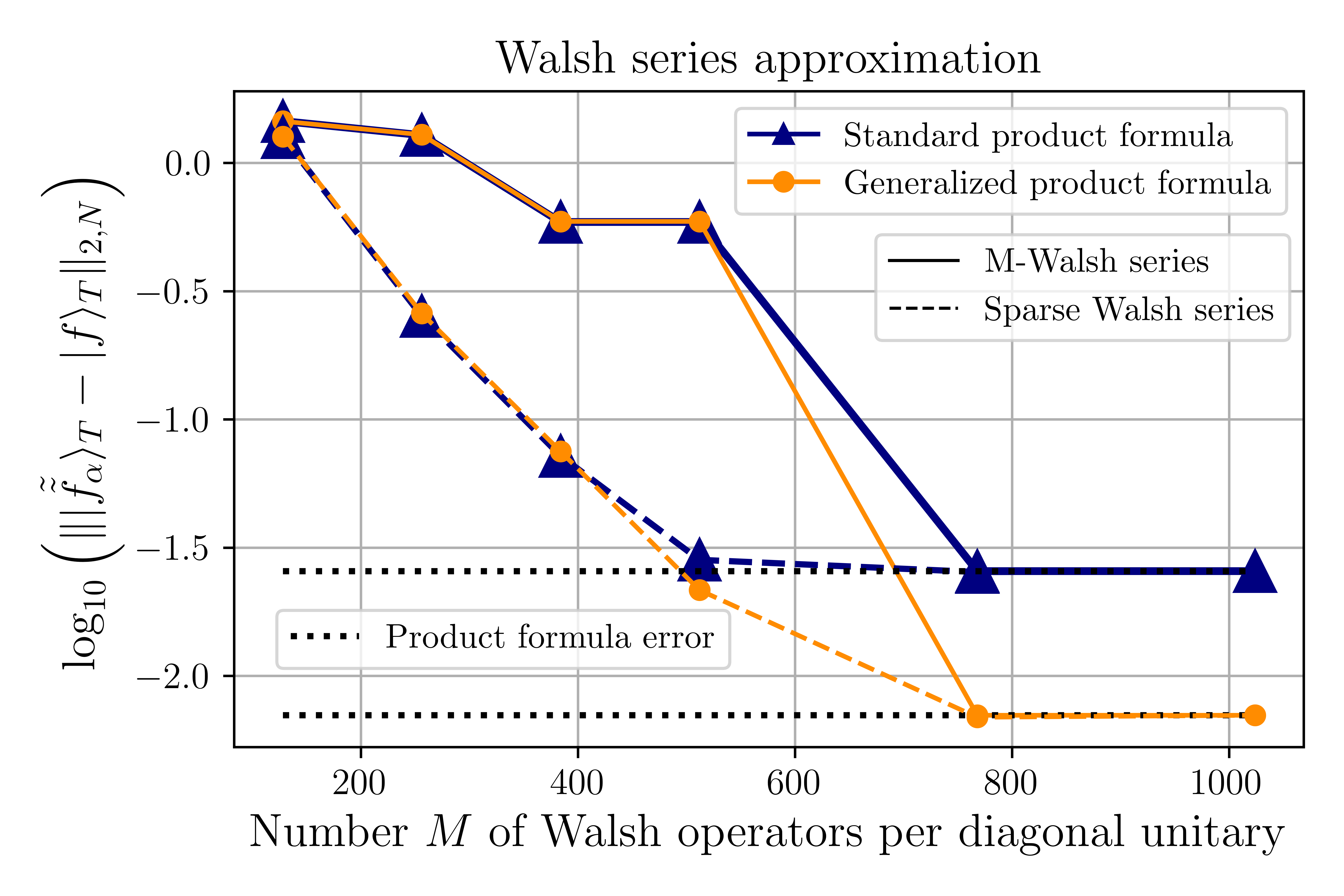}
\caption{Vector norm error $||\ket{\tilde{\tilde{f}}_{\alpha}}_T-\ket{f}_T||_{2,N}$ associated with the quantum circuit approximation as a function of the number $M$ of Walsh operators per diagonal unitaries for the generalized product formula $\alpha=g$ and the standard product formula $\alpha=s$. The number of time steps is $L=128$, the number of qubits is $n=10$, the order of the finite difference is $p=10$ and the associated two-dimensional transport equation is defined in section \ref{subsec: Collisionless Boltzmann equation} with $c_1(x_2)=(x_2-0.5), c_2(x_1)=(qE_0/m)(x_1-0.5)$ and $qE_0/m=-1$. Time for the simulation is $T=0.5$ and the initial condition is $f_0(x_1,x_2)\propto e^{-((x_1-\mu_1)^2+(x_2-\mu_2)^2)/(2\sigma^2)}$ with $\mu_1=-0.1,\mu_2=0.1$ and $\sigma=0.05$.}
\label{fig: walsh series approximation}
\end{figure}

In Figure \ref{fig: walsh series approximation}, we present the results of the quantum circuit approximation of diagonal unitaries for the two-dimensional transport equation introduced in Section \ref{subsec: Collisionless Boltzmann equation}, using $L=128$ time steps. Our numerical results show that both the $M$-Walsh series and the sparse-Walsh series (defined in appendix \ref{appendix:qc}) achieve product formula accuracy while requiring significantly fewer Walsh operators than the exact representation, which involves $1024$ Walsh operators per diagonal unitary. Moreover, the sparse-Walsh series provides a systematic tradeoff between accuracy and the number of Walsh operators, i.e., circuit size. The circuit depth could be further reduced through the use of ancilla qubits and parallelization methods for diagonal unitaries, as discussed in \cite{zylberman2025efficient}.

In the particular case where the functions $c_j$ are position-independent, one can directly parallelize the implementation of the diagonal operations and QFTs at each time-step as shown in Figure \ref{fig:schemeqc_position_independent}. Each diagonal unitary acts on a single dimension, leading to an exact complexity $C_{\text{diag,exact}}=O(2^{n_j})=O(T^{\frac{1}{2p}}d^{\frac{1}{2p}}/\epsilon^{\frac{1}{2p}})$. The size scaling remains as in Equation \ref{Eq: complexity scaling} while the depth scaling is reduced by a factor $d$ for the evolution. 

As discussed in section \ref{sec: discussion}, other methods for Hamiltonian simulation that are not based on product formula would also need to implement the diagonal operators $\hat{c}_j$ at some step of the numerical scheme, with a similar cost $O(2^n)$ for their exact implementation.

\section{Numerical simulations}
\label{sec: application}

In the following, we consider three examples of interest. The first one is the collisionless Boltzmann equation, which is a multidimensional transport equation with space- and time-dependent coefficients of the form of Eq.~\ref{Transport equation}. Then, we address the one-dimensional textbook convection equation, for which we present quantum hardware simulations. Finally, we focus on the Lotka–Volterra system, an example of nonlinear Hamiltonian ordinary differential equations that we solved via its Liouville equation, which is a transport equation of the form of Eq.~\ref{Transport equation}.

\subsection{Collisionless Boltzmann equation}
\label{subsec: Collisionless Boltzmann equation}
    Kinetic transport phenomena refer to the study of how particles, like molecules, electrons or stars, move and interact in a system when their individual motions and statistical distributions are important. These phenomena are described using kinetic theory, which provides a framework for understanding transport processes such as transport of mass, momentum or energy. For systems dominated by external forces rather than collisions, the associated equation is a linear transport equation of the form of Eq.(\ref{Transport equation}), called the collisionless Boltzmann equation:
\begin{equation}
    \partial_t f + \vec{v}\cdot\vec{\nabla}_{\vec{r}} f +\frac{\vec{F}(\vec{r},\vec{v},t)}{m}.\vec{\nabla}_{\vec{v}} f =0,
\end{equation}
where $f=f(\vec{r},\vec{v},t)$ is the probability density function of the particles defined on a six dimensional phase space $(\vec{r},\vec{v})=(x,y,z,v_x,v_y,v_z)$ and $\vec{F}(\vec{r},\vec{v},t)$ is an external time- and position-dependent force field. For instance, the probability distribution function of a population of particles of electrical charge $q$  under the action of imposed electric and magnetic fields evolves through the Lorentz force:
\begin{equation}
    \vec{F}(\vec{r},\vec{v},t)=q(\vec{E}(\vec{r},t)+\vec{v}\times \vec{B}(\vec{r},t)).
\end{equation}
In the following, we present numerical simulations of the two-dimensional collisionless Boltzmann equation with a space- and time-dependent (externally-imposed) electric field.

\subsubsection{Two-dimensional time-dependent case}
\label{subsubsec:two-dim case}

In order to illustrate the different approximations performed through the quantum numerical scheme, we consider the two dimensional collisionless Boltzmann equation with a space- and time-dependent electric field $E(x,t)=E_0(x-ct)$ for some constant $(E_0,c)\in\mathbb{R}^2$:

\begin{equation}
\partial_tf+v\partial_{x}f+\frac{qE_0}{m}(x-ct)\partial_{v}f=0,
\end{equation}
where $f$ is defined on $[0,1]^2\times[0,T]\rightarrow\mathbb{R}$ for some time $T>0$. The initial condition is:
\begin{equation}
    f(x,v,t=0)=f_0(x,v)\propto e^{-\frac{(x-\mu_1)^2}{2\sigma_1^2}-\frac{(v-\mu_2)^2}{2\sigma_2^2}},
\end{equation}
 with $(\mu_1,\mu_2,\sigma_1,\sigma_2) \in \mathbb{R}^4$\footnote{Although this initial condition is not strictly periodic, it does not introduce numerical difficulties provided the function  is negligible at the boundaries.}. 

According to the notation of Eq.(\ref{Transport equation}), this equation corresponds to a two-dimensional transport equation with $x_1=x,x_2=v$, $c_1(x_2)=x_2$, $c_2(x_1,t)=\frac{qE_0}{m}(x_1-ct)$. In this simple case, it is possible to compare the numerical solution to the analytical one, which is given by the method of characteristics as:
\begin{equation}
    f(x,v,t)=f_0(x_0(x,v,t),v_0(x,v,t)),
    \label{Eq: characteristics}
\end{equation}
where the functions $x_0,v_0$ are:
\begin{equation}
\begin{split}
    x_0(x,v,t)&=(x-ct)\cos(\omega_0t)-\frac{v-c}{\omega_0}\sin(\omega_0 t),
    \\
     v_0(x,v,t)&=c+\omega_0(x-ct)\sin(\omega_0t)+(v-c)\cos(\omega_0 t),
\end{split}
\end{equation}
with $\omega_0=\sqrt{|qE_0|/m}$.

In Figure \ref{fig: discretization error}, the error associated with the space discretization is given as a function of the number of qubits per axis for different order of the centered finite difference. Figure \ref{fig: product formula approximation} illustrates the product formula approximations as a function of the number $L$ of time steps (\ref{fig:trotter_error_with_L}), showing the expected linear decrease, and as a function of the number $n$ of qubits per axis (\ref{fig:trotter_error_with_n}), proving numerically that operator-norm bounds scale exponentially with the number of qubits per axis while the vector-norm error is independent from it. 

In kinetic theory, moments of the distribution function are of particular interest to estimate fluid quantities. For instance, the fluid density $n(x,t)=\int_v f(x,v,t) dv$ or the fluid velocity $u(x,t)=(1/n(x,t))\int_v v\,f(x,v,t)dv$ can be estimated using the observables defined in Equation \ref{eq: observables}.

\subsection{Hardware simulations of the one-dimensional transport equation}
\label{subsec:Hardware simulations of the one-dimensional transport equation}
\begin{figure}[H]
\centering
\begin{subfigure}{0.47\textwidth}
\includegraphics[width=\textwidth]{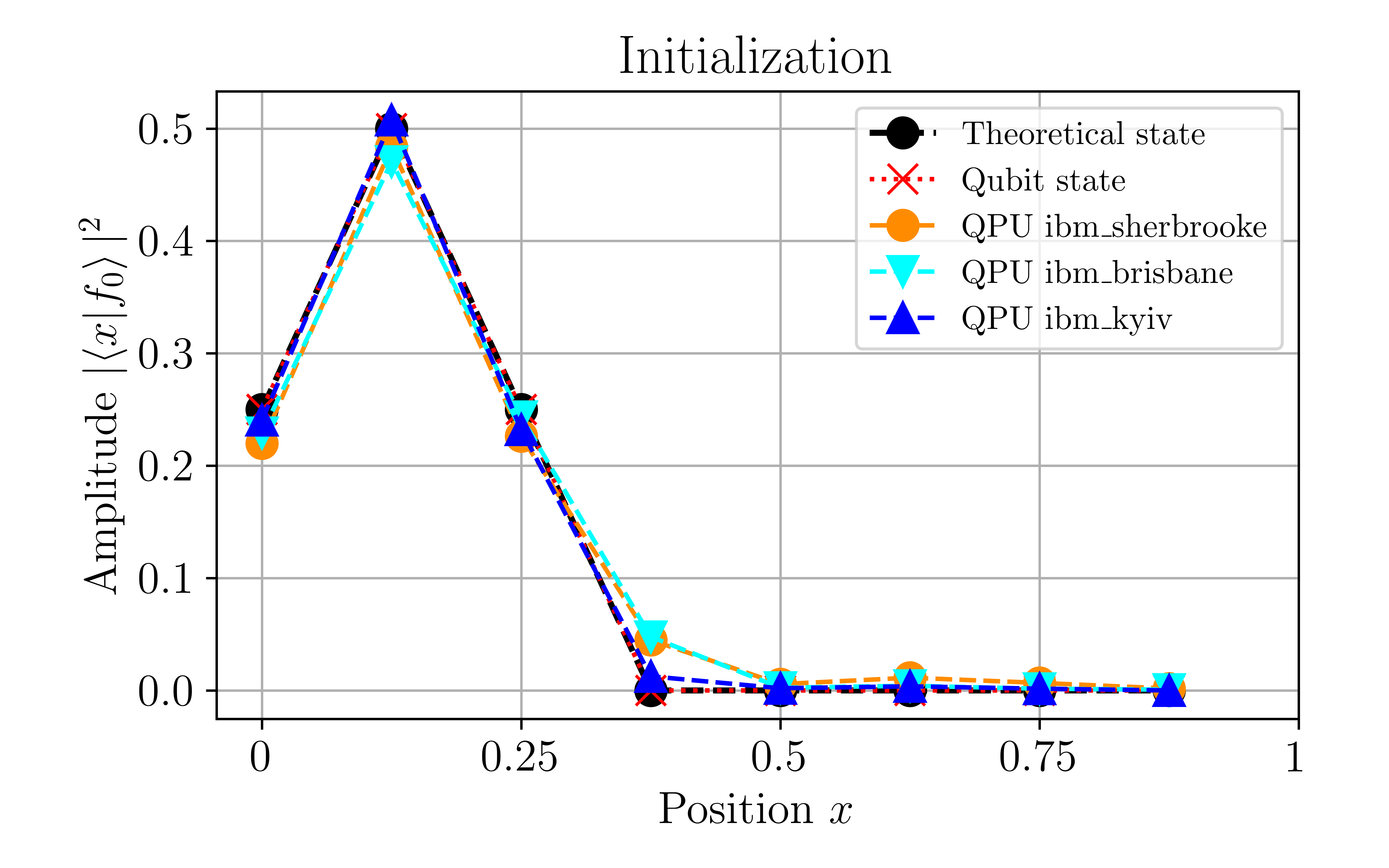}
\caption{}
\label{fig:initialisation_hardware}
\end{subfigure}
\begin{subfigure}{0.47\textwidth}
\includegraphics[width=\textwidth]{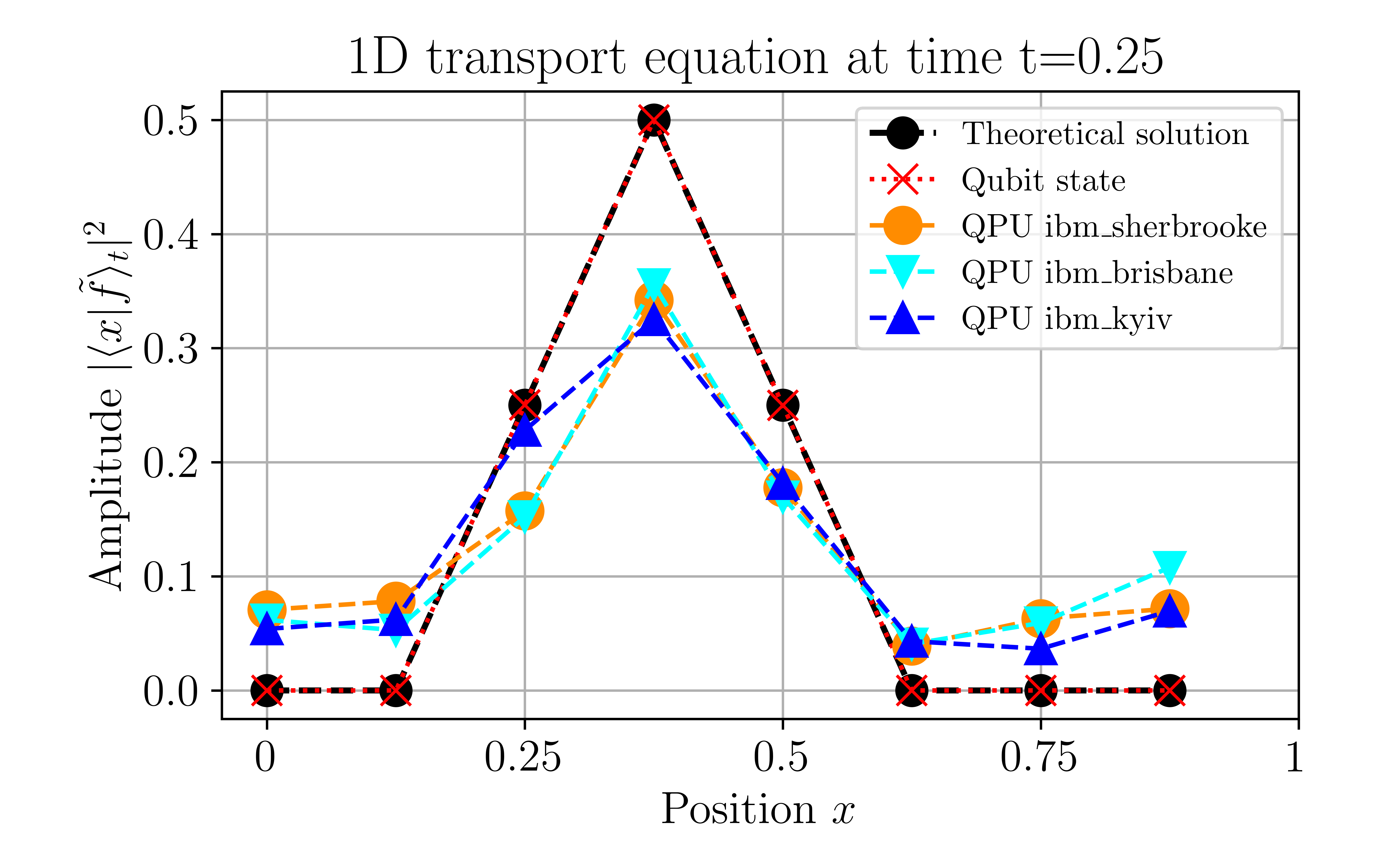}
\caption{}
\label{fig:evolution hardware t=0.25}
\end{subfigure}
\begin{subfigure}{0.47\textwidth}
\includegraphics[width=\textwidth]{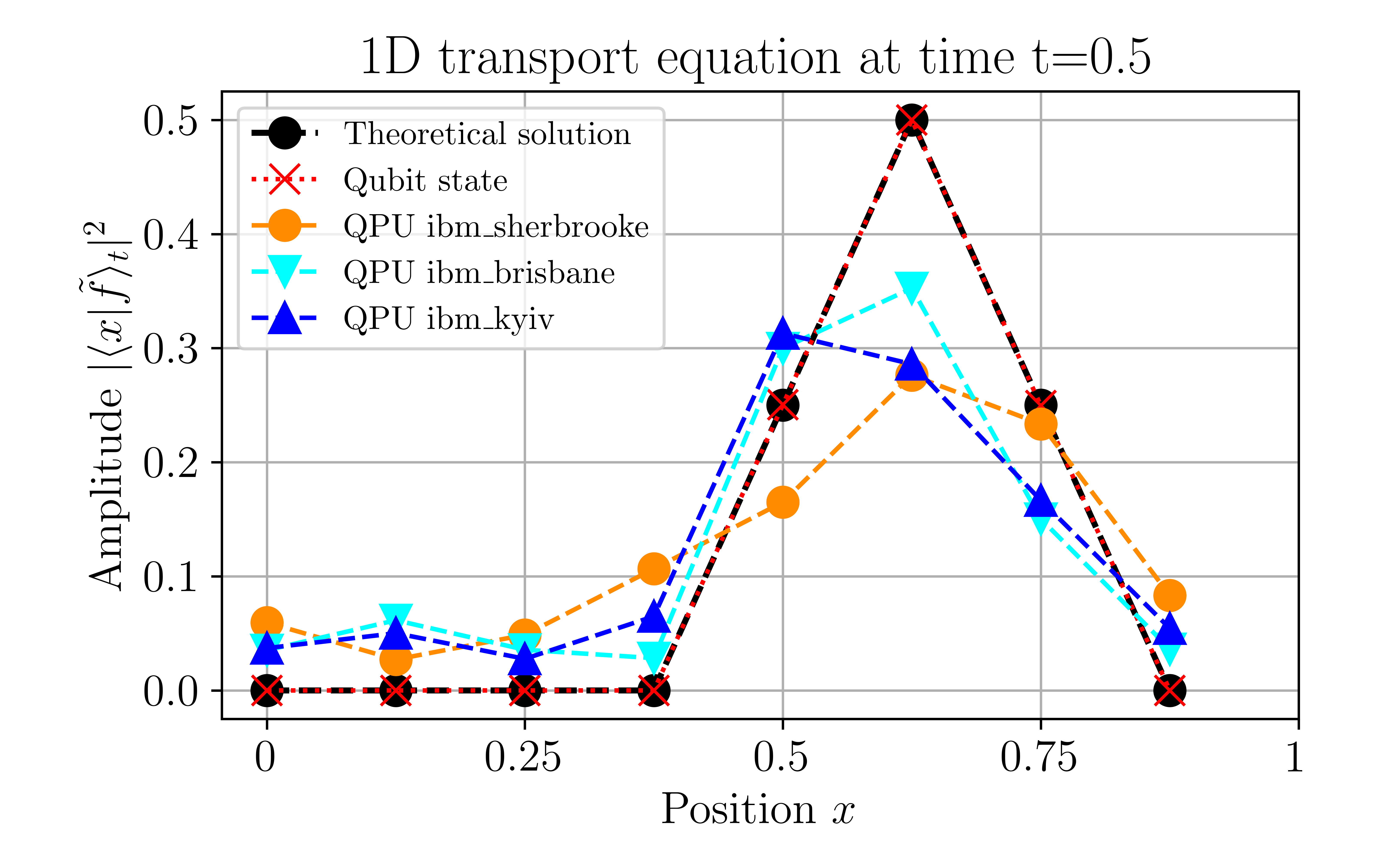}
\caption{}
\label{fig:evolution hardware t=0.5}
\end{subfigure}
\begin{subfigure}{0.47\textwidth}
\includegraphics[width=\textwidth]{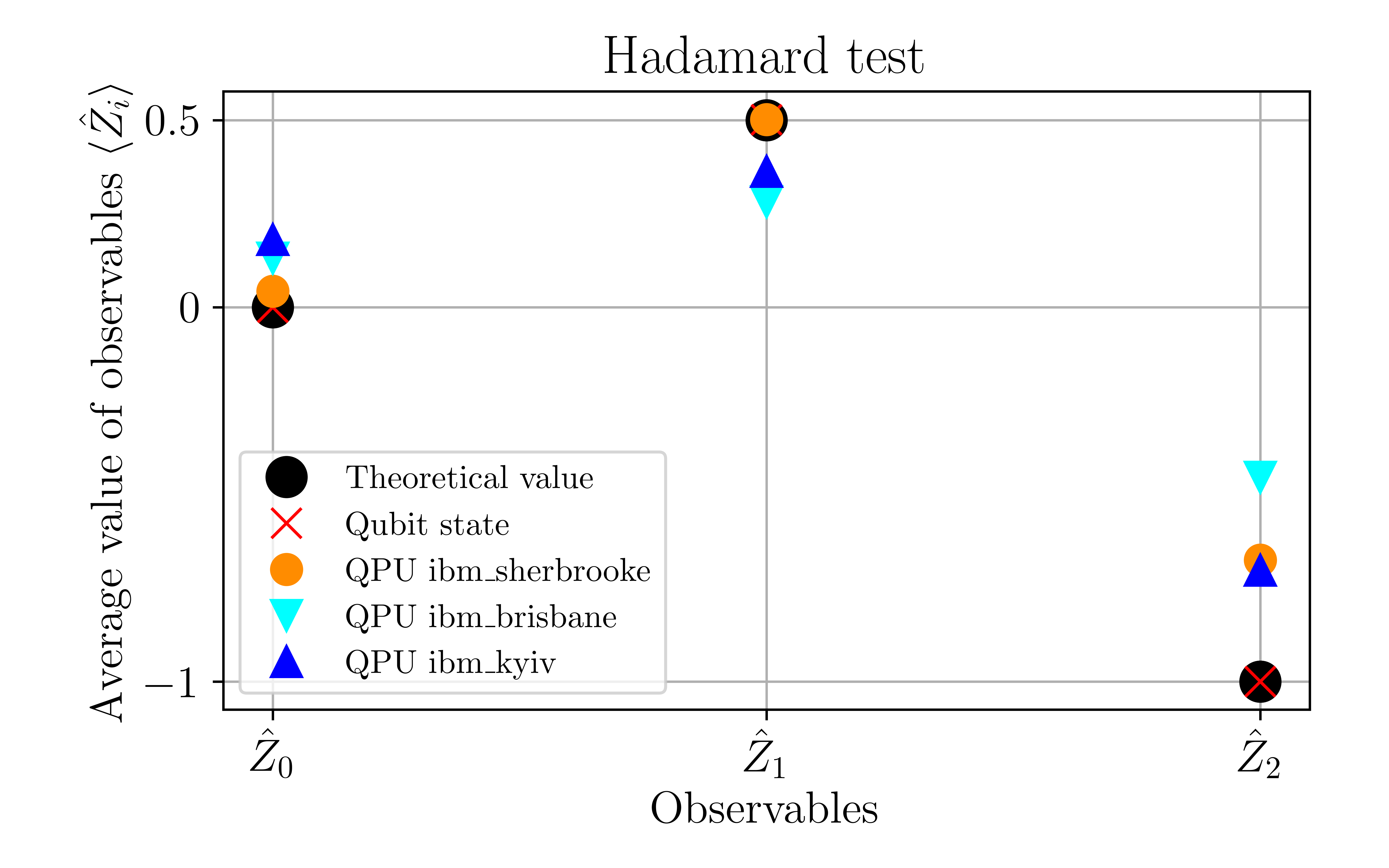}
\caption{}
\label{fig:hadamard test}
\end{subfigure}
\caption{Quantum hardware simulations of the one-dimensional convection equation defined Equation \ref{eq: one d transport equation}. The initialization and evolution are performed on $n=3$ qubits. The Hadamard test requires an additional qubit. The initial condition is given by the vector $\ket{f_0}=[1/2,1/\sqrt{2},1/2,0,0,0,0,0]$ which is implementable by using quantum gates only on two qubits among three. The evolution is performed through Equation \ref{eq: 1d transport equation ode evolution} for parameters $v=1,p=10$ at times $t=0.25$ and $t=0.5$. The quantum state preparation is exact, making the difference between the theoretical state $\ket{f_0}$ and the implemented state equal to zero. After the evolution, due to the space discretization, the difference between the theoretical solution $\ket{f}_t$ and the implemented qubit state $\ket{\tilde{f}}_t$ is $3.6\times10^{-4}$ at time $t=0.25$ and $7.2\times10^{-4}$ at time $t=0.5$. No approximation has been performed on the quantum circuits for the evolution operator. Quantum circuit sizes and depths are around $48$ and $30$ respectively for the initialization Figure (\ref{fig:initialisation_hardware}), $237$ and $154$ for the initialization and evolution in Figures \ref{fig:evolution hardware t=0.25} and \ref{fig:evolution hardware t=0.5}, and $245$ and $155$ for the initialization, evolution and Hadamard test in Figure \ref{fig:hadamard test}. }
\label{fig: quantum hardware simulations}
\end{figure}

We present real quantum-hardware simulations of the simple one-dimensional transport equation:
\begin{equation}
\begin{split}
    &\partial_tf+v\partial_xf=0,
    \\
    &f(x,t=0)=f_0(x),
\end{split}
\label{eq: one d transport equation}
\end{equation}
where $v\in\mathbb{R}^*$ is a constant velocity, $f(x,t)$ is the solution function and $f_0$ an initial condition to be specified later. After the space discretization, the evolution of the initial condition encoded unto a qubit-state is given by:
\begin{equation}
    \ket{\tilde{f}}_t=\widehat{QFT}e^{-i\frac{2vt}{\Delta x}\sum_{q=0}^{p}a_q \sin(2\pi q \hat{X})}\widehat{QFT}\ket{f_0}.
\label{eq: 1d transport equation ode evolution}
\end{equation}
Notice that in the one-dimensional case with a constant velocity $c_1=v$, there is no product formula approximation.
In Fig.~\ref{fig: quantum hardware simulations}, a $3$-qubit simulation of the one-dimensional transport equation is presented. To visualize the experiment performance, we perform the sampling of the full qubit-state at the different steps of the quantum numerical scheme. Each QPU result has required 8192 shots. The initialization, presented in Fig. \ref{fig:initialisation_hardware}, consists in preparing the qubit state $\ket{f_0}=[1/2,1/\sqrt{2},1/2,0,0,0,0,0]$ from $\ket{0}^{\otimes3}$. This is possible by applying the sequence of gates $(\widehat{SWAP}\otimes\hat{I})(C_{1\rightarrow0}(\hat{H})\otimes\hat{I})(\widehat{SWAP}\otimes\hat{I})(\hat{H}\otimes\hat{H}\otimes\hat{I})$. 
This sequence of gates has been automatically transpiled into the native set of quantum gates $\{R_Z,X,SX,ECR\}$ of the IBM's QPUs (`Eagle' processors), achieving the initialization with $48$ gates and a circuit depth of $30$\footnote{A difference of a few quantum gates has been observed from one QPU to another, likely due to other hardware constraints like connectivity}.  The results of the evolution at time $t=0.25$ are presented in Fig. \ref{fig:evolution hardware t=0.25} and, at time $t=0.5$ in Fig.~\ref{fig:evolution hardware t=0.5}.  The black curves represent the analytical solution of the transport equation \ref{eq: one d transport equation} given by $f(x,t)=f_0(x-vt)$ and the red curves are the solutions of the ODE \ref{eq: 1d transport equation ode evolution} (computed classically with the same numerical scheme). The discrepancies between the black and red curves are of order $3.6\times 10^{-4}$ at time $t=0.25$, and $7.2\times 10^{-4}$ at time $t=0.5$. They are due to the space-discretization error which grows linearly with the time $t$ (as proven in \ref{thm_discretization_error}). The expected results of the QPU are given by the red curves since no quantum circuit approximations have been performed. The observed differences are due to hardware noise (the statistical error is of order $1\%$ since $8192$ shots have been performed per experience). Despite the hardware noise, we observe the expected advection of the initial state to the positive values of $x$. A measurement protocol to estimate average value of observables has been performed with the Hadamard test, which stands as the least computationally expensive measurement protocol in terms of circuit size, depth and required ancilla qubits. Using one additional qubit, we measured the average values of the $Z$-Pauli gates applied on the $i-$th qubits $\hat{Z}_i=\ket{0}\bra{0}-\ket{1}\bra{1}$, where $i=0,1,2$.
These average values correspond to difference of the amplitudes of the qubit states as
\begin{equation}
\begin{split}
\braket{\hat{Z}_0}&=\leftindex_t{\bra{f}}(\hat{Z}\otimes\hat{I}_2\otimes\hat{I}_2)\ket{f}_t=|f_{t,0}|^2-|f_{t,0.125}|^2+|f_{t,0.25}|^2-|f_{t,0.375}|^2+|f_{t,0.5}|^2-|f_{t,0.625}|^2+|f_{t,0.75}|^2-|f_{t,0.875}|^2 , \\  \braket{\hat{Z}_1}&=\leftindex_t{\bra{f}}(\hat{I}_2\otimes\hat{Z}\otimes\hat{I}_2)\ket{f}_t=|f_{t,0}|^2+|f_{t,0.125}|^2-|f_{t,0.25}|^2+|f_{t,0.375}|^2+|f_{t,0.5}|^2-|f_{t,0.625}|^2-|f_{t,0.75}|^2-|f_{t,0.875}|^2,
\\ \braket{\hat{Z}_2}&=\leftindex_t{\bra{f}}(\hat{I_2}\otimes\hat{I}_2\otimes\hat{Z})\ket{f}_t=|f_{t,0}|^2+|f_{t,0.125}|^2+|f_{t,0.25}|^2+|f_{t,0.375}|^2-|f_{t,0.5}|^2-|f_{t,0.625}|^2-|f_{t,0.75}|^2-|f_{t,0.875}|^2,
\end{split}
\end{equation}
where $f_{t,x}$ is the component $x$ of $\ket{f}_t$. These average values are theoretically equal to $\braket{\hat{Z}_0}=0$, $\braket{\hat{Z}_1}=0.5$ and $\braket{\hat{Z}_2}=-1$ at time $t=0.5$ as displayed in Fig.~\ref{fig:hadamard test}.

\subsection{Application to non-linear ODEs: the Lotka-Volterra system}
\label{sec: non-linear ODE solver}

Some non-linear ODEs possessing a conserved quantity can be solved via a transport equation of the form of Eq.\ \ref{Transport equation}. As an example, we consider the Lotka-Volterra system: a set of non-linear ODEs that can be reformulated in Hamiltonian form. The Lotka-Volterra equations model the non-linear evolution of two interacting populations, such as predators and preys in ecology \cite{Wangerski78}, or turbulence and zonal flows in plasma physics \cite{diamond2011vorticity,ross2016predator,kobayashi2015direct,leconte2022limit}. 
The prey population, $x_1(t)\in\mathbb{R}_+^*$, grows with a rate $\alpha$ and, in presence of predators, represented by $x_2(t)\in\mathbb{R}_+^*$, decays with a rate $\beta x_2(t)$.
Similarly, the predator population grows with a rate $ \gamma x_1(t)$ and decays with a rate $\delta$.
In this representation, $\alpha,\,\beta,\,\delta,\, \gamma$ are all real and positively valued.
The non-linear ODE system describing the evolution of the populations is:

\begin{equation}
\begin{split}
\frac{dx_1(t)}{dt}&=\alpha\, x_1(t) - \beta \,x_1(t)x_2(t), \\
  \frac{dx_2(t)}{dt}&=- \delta\, x_2(t) + \gamma \, x_1(t)x_2(t).
  \label{Eq: Lotka orig}
\end{split}
\end{equation}
The previous system can be recasted in Hamiltonian form upon the change of variables $(x_1,x_2)\rightarrow(q=-\ln x_1, p=-\ln x_2)$, where the associated Hamiltonian is $H(q,p) = -\alpha\, p-\beta \,e^{-p}- \delta\, q-\gamma e^{-q}$.
In these new variables, the ODE system becomes:

\begin{equation}
\begin{split}
    \frac{dq(t)}{dt}&=-\alpha+\beta\, e^{-p(t)}\\
    \frac{dp(t)}{dt}&=\delta-\gamma \,e^{-q(t)},
\end{split}
\label{eq: Lotka voltera ODE with p,q}
\end{equation}

for which the associated Liouville equation can be written as:
\begin{equation}
  \partial_t \rho (q,p,t)=(-\delta+\gamma \,e^{-q})\,\partial_p\rho + (-\alpha +\beta \, e^{-p})\,\partial_q\rho.
\label{eq: lk transport equation}
\end{equation}
where $\rho$ denotes the probability density of the ODE solution in phase space.

\begin{figure*}[!ht]
    \includegraphics[width=\textwidth]{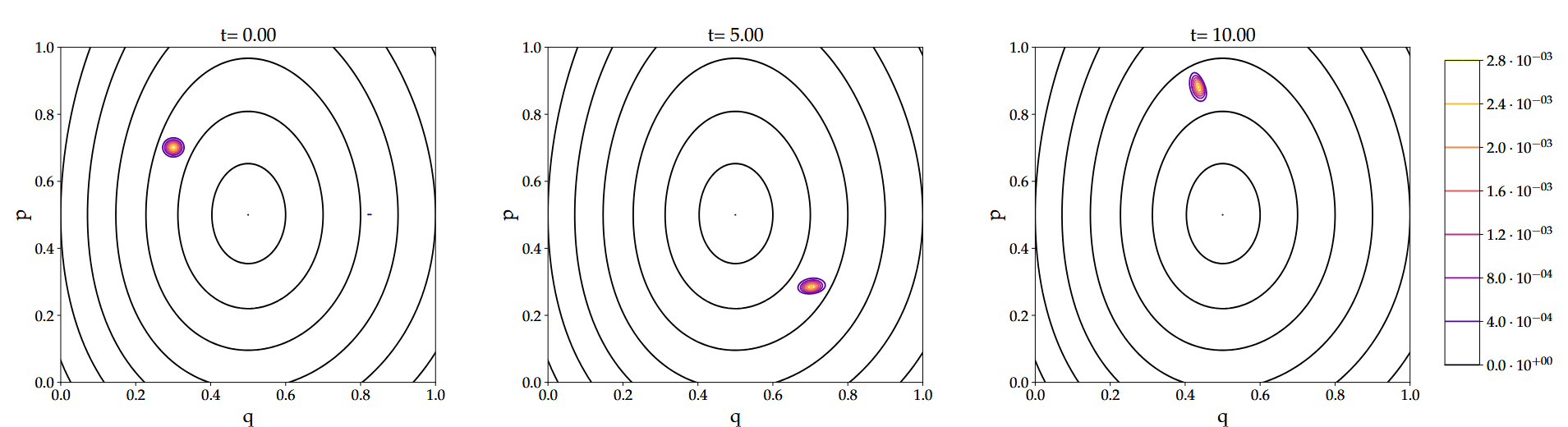}
    \caption{Evolution of the probability density associated with the Lokta-Volterra system Eq.~\ref{eq: lk transport equation}. The initial condition is $\rho_0(p,q)\propto e^{-((p-p_0)^2+(q-q_0)^2)/(2\sigma^2)}$ with $p_0=0.3, q_0=0.7$ and $\sigma=0.02$. The number of time-steps is $L=5000$ to reach time $t=5$, and $L=10000$ to reach time $t=10$. The number of qubits per axis is $n_1=n_2=9$. Parameters of the equations are $\alpha=e^{-1},\beta=e^{-1/2},\gamma=e^{1/3},\delta=e^{-1/6}$. The black lines represent different trajectories, i.e., solutions of the ODE system Eq.~\ref{Eq: Lotka orig} computed for different initial conditions.}
    \label{Fig: Lotka-volterra evolution}
\end{figure*}

The Liouville equation Eq.~\ref{eq: lk transport equation} is a transport equation of the form Eq.~\ref{Transport equation} respecting the constraint Eq.~\ref{constraint}. It can therefore be solved using the quantum numerical scheme introduced in section \ref{sec: quantum numerical scheme}.  Figure \ref{Fig: Lotka-volterra evolution} represents the evolution of an initial Gaussian probability density through the quantum numerical scheme associated with the Lotka-Volterra system\footnote{The simulation is classical, reproducing the evolution step of the quantum numerical scheme.}. Over time, the Gaussian stays localized and is only slightly distorted as it travels through phase space. The mean value of the localized probability density can be computed via the moments $\langle q\rangle(t)=\int q\,\rho(q,p,t) \,dqdp$ and $\langle p\rangle(t)=\int p\,\rho(q,p,t) \,dqdp$ and estimated via a measurement protocol presented in appendix \ref{subsec: qc for measurement}.  In Figure \ref{Fig: ODE sol}, the populations $x_1(t)=e^{-\langle q\rangle(t)}$ and $x_2(t)=e^{-\langle p\rangle(t)}$ are compared to a classical Runge-Kutta 4 solver of the ODE system defined in Eq.\ref{Eq: Lotka orig}, confirming the applicability of the approach \footnote{The averages $\langle q\rangle(t)$ and $\langle p\rangle(t)$ are computed through a classical simulation which does not include the measurement protocol and the associated errors. }.

More generally, many non-linear Hamiltonian systems can be solved via a Liouville equation in phase space \cite{hirsch2013differential,arnold2006mathematical,joseph2020koopman}, opening the possibility to study chaos, instabilities and resonances using the presented quantum numerical scheme.

\begin{figure}
    \centering
    \includegraphics[width=0.8
    \textwidth]{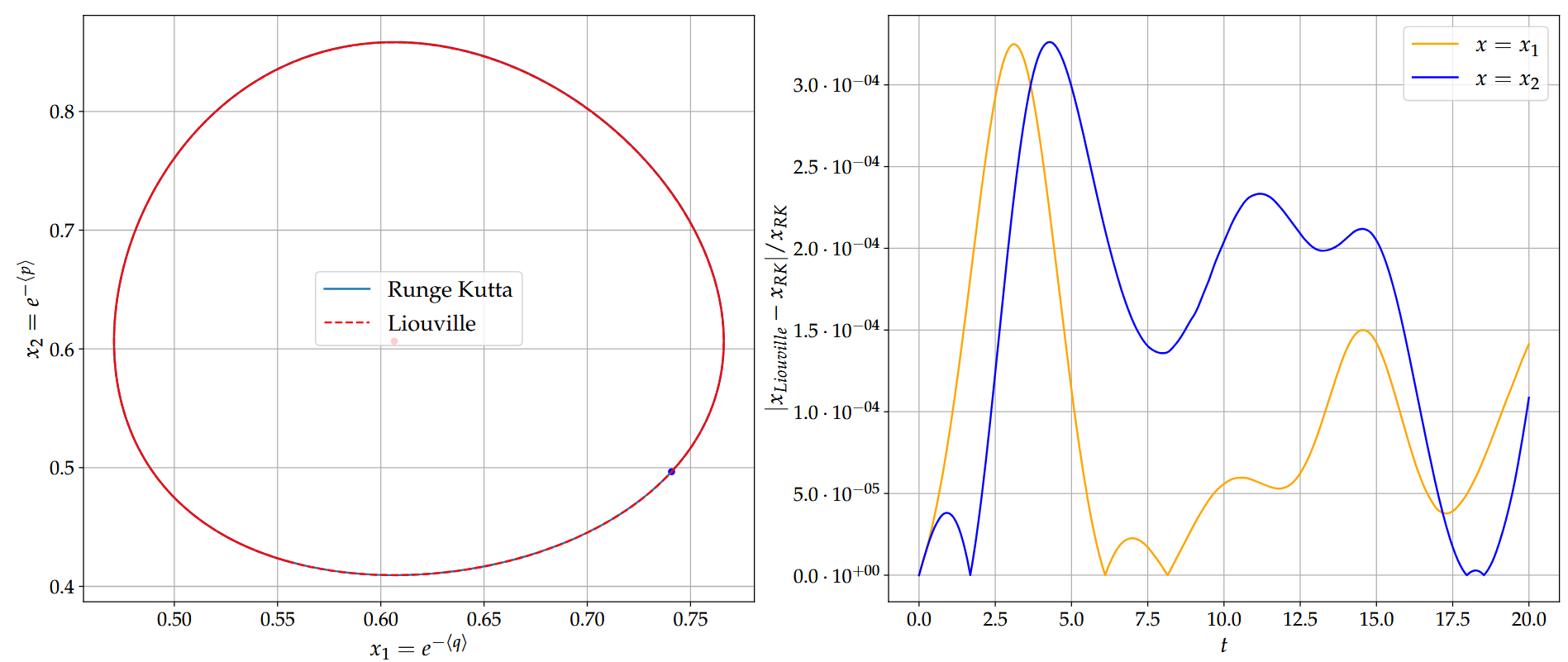}
    \caption{On the left: evolution of the moments $x_1(t)=e^{-\langle q\rangle(t)}$ and $x_2(t)=e^{-\langle p\rangle(t)}$ (red) compared with the classical Runge Kutta 4 ODE solver (blue). On the right: evolution of the relative error between $x_i(t)$ and the Runge-Kutta solutions. At the final time $T=20$, $L=2\times10^4$ time-steps have been performed.}
    \label{Fig: ODE sol}
\end{figure}

\section{Discussion and conclusion}
\label{sec: discussion}

\subsection{Comparison with prior works}

In this section, the presented quantum numerical scheme is compared to previous work.

In \cite{sato2024hamiltonian}, the authors present scalable quantum circuits to solve the $d$-dimensional transport equation with a constant vector field $\vec{c}=(c_1,\dots,c_d)\in\mathbb{R}^d$, which is time and space independent. The authors use a Matrix Product Operators (MPO) decomposition of the shift operators $\hat{S}_+=\sum_{j=0}^{2^n-1}\ket{j-1}\bra{j}=\sum_{j=1}^n s_j^+$ with $s_j^+=\hat{I}^{\otimes (n-j)}\otimes \hat{\sigma}_{01}\otimes \hat{\sigma}_{10}$ with $\hat{\sigma}_{10}=\ket{0}\bra{1}$ and $\hat{\sigma}_{10}=\ket{1}\bra{0}$, to rewrite the derivative operator as a sum of simpler operations. Then, a product formula is used to approximate the evolution operator because the $\hat{s}_j^{+}$ operators do not commute with $\hat{s}_1^{+}$ (for $j>1$).  In comparison, the scheme presented in this work addresses the problem of position and time-dependent vector field  $\vec{c}(\vec{x},t)$. In terms of circuit complexity and approximation, we directly diagonalize the derivative operators using the quantum Fourier transform, and then the diagonal unitaries are approximated using, for instance, sparse Walsh series. Even if the MPO approach could be extended to the case of space- and time-dependent $\vec{c}(\vec{x},t)$, avoiding the use of QFTs, it would still require the implementation of diagonal operators due to the operators $\hat{c}_j(t)$. Then, it is not straightforward that the numerical analysis gives better complexity scalings since the MPO decomposition requires its own product formula approximation. Additionally, their claim of scalability is based on an operator norm inequality for product formula, and it is not clear whether one can perform a vector norm analysis with the MPO decomposition.  

In \cite{hu2024quantum}, the authors use a similar MPO technique to the one introduced by \cite{sato2024hamiltonian} to solve the periodic transport equation with constant vector field $\vec{c}$. The authors use non-centered finite differences, making the evolution non-unitary. To obtain a unitary evolution, the authors introduce an additional axis which is encoded through additional ancilla qubits. Therefore, the scheme requires additional ancilla and has a probability of failure. Similarly to \cite{sato2024hamiltonian}, the analysis is based on an operator inequality and it is not clear if a vector norm analysis can be developed for MPO techniques.

In \cite{brearley2024quantum}, the authors present a quantum numerical scheme for transport equations with a time-independent vector field. The space is uniformly discretized, spatial derivatives are approximated with arbitrary finite difference, and the time derivative is approximated with a forward Euler discretization, leading to the evolution $\vec{\phi}_{t+1}=\hat{A}\vec{\phi}_{t}$ with $\hat{A}$ a non-antihermitian matrix. In order to encode $\hat{A}$, the authors introduce an ancilla qubit and prepare the operator $\exp(-i\hat{H}\theta)$ 
with $\hat{H}=\ket{0}\bra{1}\otimes (i\hat{A})+\ket{1}\bra{0}\otimes (-i\hat{A}^\dagger)$ and $\theta$ a small parameter. When $\theta$ is small enough, this operator approximates $\hat{I}_2\otimes\hat{I}+\ket{0}\bra{1}\otimes(\hat{A}\theta)+\ket{1}\bra{0}\otimes(-\hat{A}^\dagger \theta)+O(\theta^2)$. This approach approximates $\hat{A}\vec{\phi}_t$ up to some probability of success. In comparison, our scheme does not need an ancilla qubit for the evolution and we do not have a probability of success lower than $1$. In addition, our scheme is always unitary, which guarantees its stability. 

In \cite{jin2023time}, the authors propose a quantum scheme to solve nonlinear ODE based on a Liouville representation similar to what we present in section \ref{sec: non-linear ODE solver}. They consider a time-independent vector field $\vec{c}(\vec{x})$ and their approach is based on the inversion of a linear system of equations. Therefore, it does not leverage the unitarity of the evolution, but it allows to tackle non-Hamiltonian ODEs by solving the associated Koopman-Von-Neuemann equation with the quantum linear solver algorithm \cite{costa2022optimal}.

\subsection{Discussion}

The complexity of the quantum numerical scheme scales with the final time $T$ as $O(T^{2+\frac{d}{2p}})$ for an exact implementation of the diagonal unitaries. When $p$ is significantly larger than $d$, the scaling becomes $O(T^{2+o(1)})$, exceeding the optimal scaling $O(T)$ for Hamiltonian simulation algorithms \cite{berry2014exponential,berry2015simulating,low2017optimal,low2019hamiltonian}. One could consider other Hamiltonian simulation methods, based on linear combination of unitaries or Dyson series, to solve the same family of transport equations. However, the spacial dependence of the $c_j$ functions implies that, whatever the method is, one needs to ``load" the data associated with the $c_j$ in one of the steps of the quantum numerical scheme, which ultimately corresponds to implementing diagonal operations.

Despite the convergence of the numerical scheme, its efficiency is limited by the implementation of the quantum state preparation and the diagonal operators that require $O(2^n)$ quantum gates for an exact implementation. In contrast, on classical computers, initializing a vector or implementing a diagonal operation on a vector can be performed in $O(1)$ computational time since each component of the vector can be modified in parallel of the others (assuming enough parallel processors are available). However, the classical Fast Fourier transform requires exponentially more computational time  $O(N\log(N))$ than the Quantum Fourier transform $O(n^2)$, where $N=2^n$. This example illustrates that limiting tasks on quantum computers can be relatively simple on a classical one, and vice-versa. However, in cases where the initial condition and the diagonal operators are efficiently implementable, it may be possible that quantum computers can estimate quantities of interest from PDEs solutions more efficiently than classical computers. 

Furthermore, removing the assumption Eq.\ref{constraint}, which states that each $c_j$ function is independent of the $j$-th variable $x_j$,  requires the modification of the numerical scheme since it implies that the evolution operator becomes non-unitary. For instance, in the one-dimensional (time-independent) case, one obtains an evolution operators of the form $e^{-ic(\hat{x})\hat{D}_xt}$ that is not diagonalized by the quantum Fourier transform. This operator could be implemented using an other type of quantum numerical scheme, for instance via a linear combination of Hamiltonian simulations (LCHS) ~\cite{an2023linear,an2024laplace}. However, it remains challenging to implement the obtained unitary operators via the LCHS, due to the fact that the associated Hamiltonian are non-sparse operators that are non-trivially diagonalizable. Additionally, performing a vector norm analysis could be challenging since one would need to take into account the dissipation and the amplification of modes of the solution function induced by the non-unitary evolution.

\subsection{Conclusion}
In this work, we present a quantum numerical scheme for the multidimensional transport equation with space- and time-dependent coefficients based on three stages: initialization, evolution, and measurement. The evolution step employs high-order finite-difference approximations of spatial derivatives, combined with product-formula (Trotterization) approximations of the evolution operator. Our novel numerical analysis improves existing operator norm inequalities by a factor $\Theta(4^{n})$, thereby guaranteeing convergence with a number of time-steps reduced by an exponential-with-$n$ factor. Simulations of a two-dimensional transport equation demonstrate the different error scalings and illustrate the effectiveness of sparse Walsh series in reducing circuit size and depth. Pedagogical simulations of the canonical one-dimensional convection equation were performed on noisy quantum hardware to validate the approach. Furthermore, we demonstrate the potential of this framework for solving nonlinear Hamiltonian ODEs via their Liouville formulation, exemplified by the Lotka–Volterra system.

These results provide a robust and practical framework for simulating transport phenomena on quantum computers.  Future work could focus on developing a vector-norm analysis for quantum algorithms beyond Trotterization, as well as for PDE problems involving dissipation, stochasticity, or non-linearities. 

\section{Acknowledgments}
The authors would like to thank A.~Ameri, M.~Arsenault, L.~Boudin, M.~Brachet, B.~Claudon, C.~Feniou, T.~Laakonen, L.~Londeix Pagnard and U.~Nzongani for useful discussions on the content of this work. T. Fredon and N. F. Loureiro were funded by U.S. Department of Energy Grant No. DE-SC0020264 and by the MIT-IBM Watson AI Lab. We acknowledge the use of IBM Quantum services for this work. The views expressed are those of the authors, and do not reflect the official policy or position of IBM or the IBM Quantum team. We also acknowledge the use of Qiskit \cite{qiskit2024} for the design and implementation of the quantum circuits.

\bibliography{main.bib}
\appendix

\section{Quantum circuit constructions}
\label{appendix:qc}

In this section, we provide quantum circuit constructions associated with the different steps of the quantum numerical scheme.

\subsection{Quantum state preparation}
\label{Appendix: quantum state preparation}

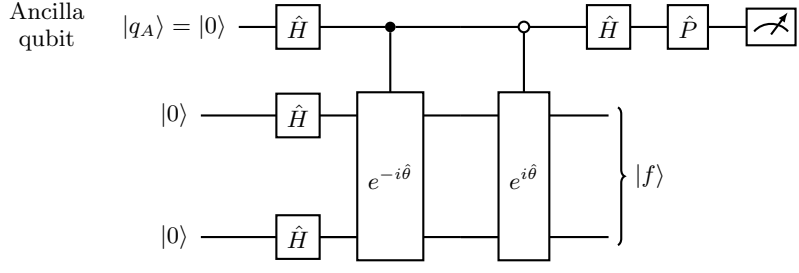
\begin{figure}[ht]
    \centering
\begin{quantikz}
\begin{matrix} \text{Ancilla} \\ \text{qubit} \end{matrix}& \ket{q_A}=\ket{0} \text{ }& \gate{\hat{H}} 
&\ctrl{1} &\qw& \octrl{1}& \gate{\hat{H}}& \gate{\hat{P}} & \meter{} 
\\
&\ket{0} \text{ }&
     \gate{\hat{H}}
&\gate[3,nwires={2}]{e^{-i\hat{\theta}}} & \qw &\gate[3,nwires={2}]{e^{i\hat{\theta}}}& \qw \rstick[wires=3]{$\ket{f}$} &&
\\ & & && &&&&&&
\\& \ket{0} \text{ }&\gate{\hat{H}} & \qw &\qw&\qw&\qw&& 
\end{quantikz} 
\caption{Quantum state preparation \cite{zylberman2025efficient} for qubit states $\ket{f}$ with real-valued components using two controlled-diagonal unitary. The operator $\hat{\theta}$ encodes the components of $\ket{f}$ as $\hat{\theta}=\sum_{\vec{X}}\arcsin(f(\vec{X})/(\alpha f_{\max}))\ket{\vec{X}}\bra{\vec{X}}$ with $\alpha\ge 1$. The gate $\hat{P}=\begin{pmatrix}1&0\\0&-i\end{pmatrix}$ is optional.}
\label{fig:quantum circuit scheme for qsp two dia}
\end{figure}

Many protocols for quantum state preparation have been developed in the past few years \cite{sun2023asymptotically,zhang2022quantum,yuan2023optimal,zylberman2024efficient,zylberman2025efficient,moosa2023linear}. For completeness of this work, we briefly present an example of protocol from reference \cite{zylberman2025efficient}, which uses one ancilla qubit and two controlled-diagonal unitaries to efficiently implement qubit states depending on differentiable functions, as it is the case for the initialization of the PDE problem. The protocol, presented in Figure \ref{fig:quantum circuit scheme for qsp two dia}, uses two controlled-diagonal unitaries, one being controlled by the ancilla qubit in state $\ket{1}$ and the other being "anti-controlled", i.e. controlled by the ancilla qubit in state $\ket{0}$. Each diagonal unitary is the Hermitian conjugate of the other. The core idea is to use one ancilla to create the operator $-i(e^{i\hat{\theta}}-e^{-i\hat{\theta}})/2=\sin(\hat{\theta})$. Then, one can prepare the target quantum state up to some probability of success by choosing the phase as $\hat{\theta}=\arcsin(\hat{f}/(\alpha f_{\max}))$, where $\alpha \ge 1$ is an optional parameter and $f_{\max}=\max_{\vec{X}}|f(\vec{X})|$. The qubit state $\ket{0}^{\otimes(n+1)}$ evolves through the quantum circuit represented in Figure \ref{fig:quantum circuit scheme for qsp two dia} as:
\begin{equation}
\begin{split}
    \ket{0}^{\otimes n}\ket{0}&\xrightarrow[]{\hat{H}^{\otimes(n+1)}}\frac{1}{\sqrt{2}}(\ket{s}\ket{0}+\ket{s}\ket{1})\\
    &\xrightarrow[]{C^1(e^{-i\hat{\theta}})} \frac{1}{\sqrt{2}}(\ket{s}\ket{0}+e^{-i\hat{\theta}}\ket{s}\ket{1}) \\
    &\xrightarrow[]{\overline{C^1}(e^{i\hat{\theta}})} \frac{1}{\sqrt{2}}(e^{i\hat{\theta}}\ket{s}\ket{0}+e^{-i\hat{\theta}}\ket{s}\ket{1}) \\
    &\xrightarrow[]{\hat{I}\otimes(\hat{P}\hat{H})}
    \frac{e^{i\hat{\theta}}+e^{-i\hat{\theta}}}{2}\ket{s}\ket{0}-i\frac{e^{i\hat{\theta}}-e^{-i\hat{\theta}}}{2}\ket{s}\ket{1}\\
    &=\cos(\theta)\ket{s}\ket{0}+\sin(\theta)\ket{s}\ket{1}\\
    &\xrightarrow[]{\text{If } \ket{q_A}=\ket{1}:}
    \frac{\sin(\hat{\theta})}{\|\sin{(\hat{\theta})}\ket{s}\|_{2,N}}\ket{s}=\ket{f},
\end{split}
\label{eq:realqsptwodiag}
\end{equation}
where $\overline{C^1}(\hat{U})=(\hat{X}\otimes \hat{I}_T)C^1(\hat{U})(\hat{X}\otimes \hat{I}_T)$ is the anti-controlled $\hat{U}$ operation, where $\hat{X}$ acts on the qubit of control and $\hat{I}_T$ is the identity on the target position register. 

This quantum state preparation succeed when the ancilla qubit is measured in state $\ket{1}$ with a probability of success:
\begin{equation}
    P(1)=\|\sin(\hat{\theta})\ket{s}\|_{2,N}^2=\|\frac{\hat{f}\ket{s}}{\alpha f_{\max}}\|_{2,N}^2=\frac{1}{\alpha^2 N}\sum_{\vec{X}} \frac{|f(\vec{X})|^2}{f_{\max}^2}\xrightarrow[]{N\rightarrow +\infty}\frac{\|f\|_{L^2}^2}{\alpha^2 \|f\|_\infty^2}=\Theta(1),
\end{equation}
where $\|f\|_{L^2}=\sqrt{\int_{[0,1]^d}|f(\vec{X})|^2dV}$ and $\|f\|_{\infty}=\max_{x\in[0,1]}|f(x)|$. Notice that one can use amplitude amplification to make the probability of success closer to $1$, or perform the protocol until a success is achieved. The parameter $\alpha \ge 1$ is optional but necessary for Walsh series analysis. As explained in the following section, diagonal unitaries depending on differentiable functions can be implemented through a Walsh series approximation, which depends on the maximum value of the derivative and is bounded for the function $g(x)=\arcsin(f(x)/(\alpha f_{\max}))$ when $\alpha>1$.

\subsection{Quantum circuits for diagonal unitaries}
\label{subsec: qc for diagonal unitaries}

Different methods exist to implement diagonal unitaries based either on a sequential decomposition, a Walsh-Hadamard decomposition or a Fourier decomposition. In the following, we present the quantum circuits associated with the Walsh-Hadamard decomposition, which are used to implement the diagonal unitaries $\hat{\Lambda}_{j,\alpha}$ defined in Equation \ref{eq: diag unitaries}, which depend on differentiable functions. Additional details can be found in reference \cite{zylberman2025efficient}. First, we recall elementary results on Walsh functions and Walsh series. Then, we introduce the associated quantum circuits for diagonal unitaries.

\subsubsection{Walsh functions, Walsh operators and Walsh series}

\paragraph{Walsh functions and operators.}
Walsh functions were developed by Joseph L.Walsh in 1923 \cite{walsh1923closed}, who showed that every continuous function of bounded variation defined on $[0,1]$ can be expanded into a series of Walsh functions. Walsh functions form a set of orthogonal functions defined on $[0,1]$ by
\begin{equation}
    w_j(x)=(-1)^{\sum_{j=1}^nj_ix_{i-1}},
\end{equation}
where $j$ is the order of the Walsh function, $j_i$ is the $i$-th bit in the binary expansion $j=\sum_{i=1}^nj_i2^{i-1}$ and $x_i$ is the $i$-th bit in the diadic expansion $x=\sum_{i=0}^{\infty}x_i/2^{i+1}\in[0,1]$. Walsh functions take only two different values $\pm1$ , which make them natural to implement on digital computers where binary logic is at the core of the routines. Indeed, the Walsh operator $\hat{w}_j$ associated with the Walsh function $w_j$ as $\hat{w}_j=\sum_{x\in\mathcal{X}_n}w_j(x)\ket{x}\bra{x}$ is equal to a tensor product of $\hat{Z}=\ket{0}\bra{0}-\ket{1}\bra{1}$ Pauli  gates as
\begin{equation}
    \hat{w}_j=(\hat{Z}_1)^{j_1}\otimes\hdots\otimes (\hat{Z}_n)^{j_n},
\label{Eq: walshoperator}
\end{equation}
where $j_i$ is the $i$-th coefficient in the binary decomposition of $j=\sum_{i=1}^nj_i2^{i-1}$.

\paragraph{Walsh series.}
Walsh series can represent functions $f$ defined on $[0,1]$ with bounded variations. On $[0,1]$, the $M$ Walsh series of $f$ is defined as:
\begin{equation}
    S_{f,M}=\sum_{j=0}^Ma_j^fw_j,
\end{equation}
where $a_j^f$ is the $j$-th Walsh coefficients of $f$ defined by:
\begin{equation}
    a_j^f=\frac{1}{M}\sum_{k=0}^{M-1}f(k/M)w_j(k/M).
\end{equation}
On $\mathcal{X}_m=\{0,1/M,\hdots,(M-1)/M\}$, the $(M=2^m)$-Walsh series of $f$ is an exact representation of $f$: $\forall x \in \mathcal{X}_m$, $f(x)=S_{f,M}(x)$. On $[0,1]$, the $M$-Walsh series of $f$ approximates $f$ up to an error that decreases linearly with the number of terms $M$:
\begin{equation}
    \|f-S_{f,M}\|_{\infty}\leq \frac{\|f'\|_{\infty}}{M},
\end{equation}
where $\|f'\|_{\infty}=\max_{x\in[0,1]}|f'(x)|$. A proof of this inequality can be found in the Lemma 2 of the appendix B.2.a. of reference \cite{zylberman2024efficient}.
By choosing $M=\lceil \|f'\|_{\infty}/\epsilon \rceil$, one gets an $(\epsilon>0)$-approximation of $f$ in terms of Walsh series.

\paragraph{Sparse Walsh series.}

In references \cite{zylberman2024efficient} and \cite{zylberman2025efficient}, sparse Walsh series were introduced as $f_s=\sum_{j\in S}a_jw_j$ with $S\subseteq \{0,1,\hdots,2^n-1\}$ and the $a_j$ coefficients can be chosen to minimize the difference between $f_s$ and $f$. The problem of finding the best set $S$ and the best coefficients $\{a_j, j\in S\}$ is called the minimax problem \cite{yuen1975function}. A simple but efficient way to find a sparse Walsh series approximating a given function $f$ is to sort the Walsh coefficients of the $M$-Walsh series and to keep the largest coefficients $|a_j|$. In Figure \ref{fig: walsh series approximation}, we compare sparse Walsh series and $M$-Walsh series for approximating the different diagonal unitaries. In this example, we numerically prove that sparse Walsh series require less terms than $M$-Walsh series to achieve the same approximation level.

\paragraph{Multidimensional Walsh series.}
The approximation of functions by Walsh series can be extended to the multi-variate case as:
\begin{equation}
    S_{f,\vec{M}}=\sum_{j_1=0}^{M_1-1}\hdots\sum_{j_d=0}^{M_d-1}a_{j_1,\hdots,j_d,M_1,\hdots,M_d}^f(\vec{r})=\sum_{\vec{j}}a_{\vec{j},\vec{M}}^f w_{\vec{j}}(\vec{r}),
\end{equation}
where $w_{j_1,\hdots,j_d}=w_{\vec{j}}(\vec{r})=w_{j_1}(r_1)\times \hdots \times w_{j_d}(r_d)$ and $a_{\vec{j},\vec{M}}^f$ is the multidimensional Walsh coefficient:
\begin{equation}
    a_{\vec{j},\vec{M}}^f=\frac{1}{M}\sum_{\vec{r}\in\mathcal{X}_{\vec{m}}}f(\vec{r})w_{\vec{j}}(\vec{r}),
\end{equation}
where $M=M_1\times \hdots \times M_d=2^{m_1}\times  \hdots \times 2^{m_d}$.

For any differentiable function $f$ defined on $[0,1]^d$, the $\vec{M}=(M_1,\hdots,M_d)$-Walsh series $S_{f,\vec{M}}$ approximates $f$ on $[0,1]^d$ as:
\begin{equation}
    \|f-S_{f,\vec{M}}\|_{\infty}\leq\sum_{i=1}^d\frac{\|\partial_{x_i}f\|_{\infty}}{M_i}.
\end{equation}
The proof is given in Lemma 9 of appendix B.9.a. of publication \cite{zylberman2024efficient}. One can design a Walsh series $S_{f,\vec{M}}$ that is an $\epsilon>0$ approximation of $f$ by choosing $M_i=\lceil \|\partial_{x_i}f\|_{\infty}/\epsilon \rceil $, $\forall i\in\{1,\hdots,d\}$. The $\epsilon-$approximation $S_{f,\vec{M}}$ contains $M_1\times \hdots \times M_d=O(1/\epsilon^d)$ terms.

\subsubsection{Walsh-Hadamard decompositions of diagonal unitaries}

We introduce the quantum circuits for implementing a diagonal unitary $e^{i\hat{f}}$ acting on $n=n_1+\hdots+n_d$ qubits and depending on a multivariate differentiable function $f$ that is approximated with a Walsh series $S_{f}=\sum_{\vec{j}\in S}a_{\vec{j}}w_{\vec{j}}$, where $S\subseteq \{0,1,\hdots,2^{n_1}-1\}\times\hdots\times\{0,1,\hdots,2^{n_d}-1\}$. First, notice that if $S_f$ is an $\epsilon$-approximation of $f$ as $\|f-S_f\|_{\infty}\leq\epsilon$, then $e^{i\hat{S}_f}$ is an $\epsilon$-approximation of $e^{i\hat{f}}$ in spectral norm as $\|e^{i\hat{f}}-e^{i\hat{S}_f}\|_2\leq \epsilon$. Then, the implementation of $e^{i\hat{S}_f}$ is simply given by the product of the exponential of the Walsh operators thanks to the fact that each Walsh operator $\hat{w}_{\vec{j}}$ commutes with the other:
\begin{equation}
    e^{i\hat{S}_f}=\prod_{\vec{j}\in S}e^{ia_{\vec{j}}\text{ }\hat{w}_{\vec{j}}}.
\end{equation}

\begin{figure}[h]
    \centering
\begin{quantikz}
\lstick[wires=1]{$q_0$} 
&\gate[1]{\hat{Z}} & \qw & && \qw & \ctrl{4} & \qw & \qw& \qw&\qw & \qw&\qw & \qw & \ctrl{4}&\qw
\\ 
 \lstick[wires=1]{$q_1$} 
&\gate[1]{\hat{Z}} & \qw & && \qw & \qw & \ctrl{3}& \qw& \qw &\qw &\qw & \qw &\ctrl{3} & \qw& \qw
\\
& \vdots & & \equiv & &&&& \ddots && && \iddots&
\\  \lstick[wires=1]{$q_{p-2}$} & \gate[1]{\hat{Z}} & \qw &  & & \qw & \qw & \qw& \qw  & \ctrl{1} & \qw & \ctrl{1} & \qw& \qw & \qw & \qw
\\
\lstick[wires=1]{$q_{p-1}$}& \gate[1]{\hat{Z}}& \qw & && \qw & \targ{} &\targ{}& \qw & \targ{} & \gate[1]{\hat{Z}} & \targ{} & \qw&\targ{} &\targ{} & \qw
\\
\end{quantikz}
\caption{ Equivalent quantum circuit for a tower of $p$ Z-Pauli quantum gates. }
\label{fig:Z tower}
\end{figure}
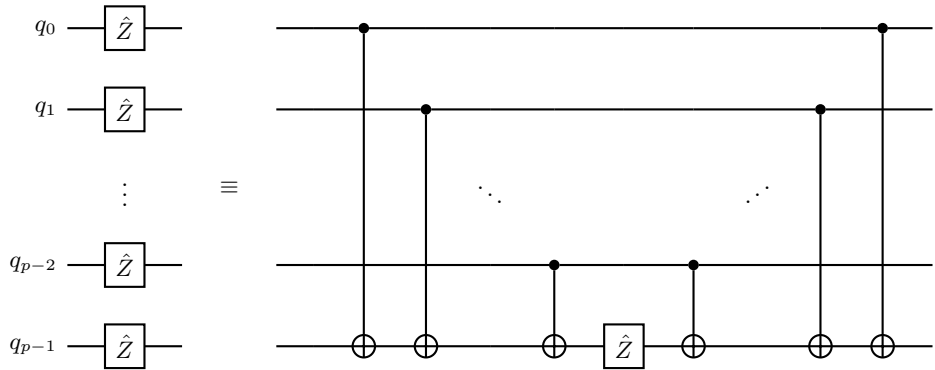

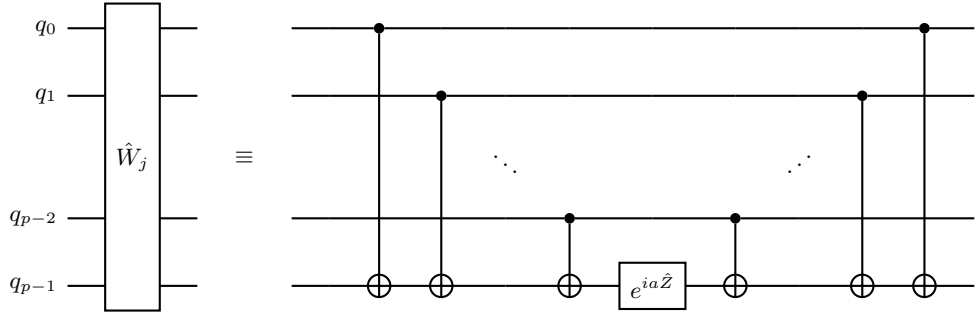
\begin{figure}[h]
    \centering
\begin{quantikz}
\lstick[wires=1]{$q_0$} 
&\gate[5,nwires={3}]{\hat{W}_{j}} & \qw & && \qw & \ctrl{4} & \qw & \qw& \qw&\qw & \qw&\qw & \qw & \ctrl{4}&\qw
\\ 
 \lstick[wires=1]{$q_1$} 
& & \qw & && \qw & \qw & \ctrl{3}& \qw& \qw &\qw &\qw & \qw &\ctrl{3} & \qw& \qw
\\
& \vdots & & \equiv & &&&& \ddots && && \iddots&
\\  \lstick[wires=1]{$q_{p-2}$} &  & \qw &  & & \qw & \qw & \qw& \qw  & \ctrl{1} & \qw & \ctrl{1} & \qw& \qw & \qw & \qw
\\
\lstick[wires=1]{$q_{p-1}$}& & \qw & && \qw & \targ{} &\targ{}& \qw & \targ{} & \gate[1]{e^{ia\hat{Z}}} & \targ{} & \qw&\targ{} &\targ{} & \qw
\\
\end{quantikz}
\caption{Quantum circuit for the operator $\hat{W}_j$ acting on $p$ different qubits using CNOT and RZ quantum gates.}
\label{fig:Z tower2}
\end{figure}

Using the fact that a tensor product of $\hat{Z}$-Pauli gates can be rewritten using two CNOT stairs and one $\hat{Z}$-Pauli gate, as represented in Figure \ref{fig:Z tower}, the operators $\hat{W}_{\vec{j}}=e^{ia_{\vec{j}}\hat{w}_{\vec{j}}}$ can be implemented using one $\hat{R}_Z$ gate and $2k_{\vec{j}}$ CNOT gates, where $k_{\vec{j}}=H(j_1)+\hdots+H(j_n)\leq n$ is the sum of the Hamming weight of each index $j_{i}=\sum_{l=1}^nj_{i,l}2^{l-1}$ with $j_{i,l}\in \{0,1\}$ and $H(j_i)=\sum_{l=1}^nj_{i,l}\leq n_i$.
For example, for $j=\sum_{i=0}^{p-1}j_i2^{i}$ with $1\leq p\leq n$ and $j_{p-1}=1$, the operator $\hat{W}_j$, represented in Figure \ref{fig:Z tower2}, acts on the $p$-th qubits $0,...,p-1$:
\begin{equation}
   \hat{W}_j=\left(\prod_{i=0}^{p-2}\widehat{CNOT}_{i\rightarrow p-1}^{j_i}\right)\left(\hat{I}_2^{\otimes(p-1)} \otimes \hat{R}_Z(a_j))
   \right)\left(\prod_{i=0}^{p-2}\widehat{CNOT}_{i\rightarrow p-1}^{j_i}\right)^{-1},
\label{eq: exp of walsh operator}
\end{equation}
where $\widehat{CNOT}_{i\rightarrow p-1}^{j_i}$ is the CNOT gate controlled by the $i$-th qubit and applied on the $(p-1)$-th qubit if $j_i=1$, the symbol $\prod_{i=0}^{p-2}\hat{A}_i=\hat{A}_0...\hat{A}_{p-2}$ is the product of operator $\hat{A}_i$ with indexes in increasing order, $\hat{I}_2^{\otimes (p-1)}$ is the identity operator on the qubits $0$ to $p-2$ and $\hat{R}_Z(a_j)=e^{i a_j}\ket{0}\bra{0}+e^{-i a_j}\ket{1}\bra{1}$ acts on the qubit $p-1$. The operator $\hat{W}_0(a_0)=e^{ia_0}\hat{I}_2^{\otimes n}$ is a phase encoding the average value of the $\{\theta_k\}$.

Additionally, remark that the Walsh coefficient of order zero in the Walsh series of $f$ corresponds to its average value. Its implementation corresponds to a global phase for the non-controlled diagonal unitary which becomes a relative phase, i.e. a non-optional two-qubit gate, when the diagonal unitary is controlled.

\subsection{Quantum circuits for measurement protocols}
\label{subsec: qc for measurement}

\subsubsection{Hadamard test}

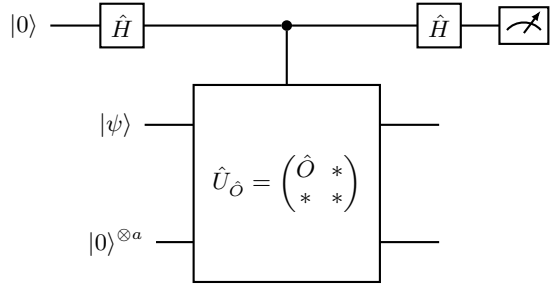
\begin{figure}[ht]
    \centering
\begin{quantikz}
 \ket{0}\text{ } &\gate{\hat{H}} 
&\ctrl{1} & \gate{\hat{H}}&  \meter{} 
\\
&\ket{\psi}\text{ } &\gate[2,nwires={2}]{\hat{U}_{\hat{O}}=\begin{pmatrix}
    \hat{O} & * \\ * & *
\end{pmatrix}} & \qw
\\ &\ket{0}^{\otimes a}\text{ } & \qw &\qw
\end{quantikz} 
\caption{Hadamard test for estimating $\Re(\bra{\psi}\hat{O}\ket{\psi})$ where $\hat{U}_{\hat{O}}$ is a block-encoding of an operator $\hat{O}$, with $\|\hat{O}\|_2\leq 1$ using $a$ ancilla qubits. The quantity $\Im(\bra{\psi}\hat{O}\ket{\psi})$ can be estimated with the same protocol using an additional phase gate $\hat{P}=\ket{0}\bra{0}+i\ket{1}\bra{1}$ on the first ancilla qubit after the first Hadamard gate.}
\label{fig: Hadamard test circuit}
\end{figure}

The Hadamard test presented in Figure \ref{fig: Hadamard test circuit} is one of the simplest measurement protocol. It makes use of one ancilla qubit to control the application of an operator $\hat{U}_{\hat{O}}$ that block-encodes an operator $\hat{O}$\footnote{When $\hat{O}$ is unitary, they are no specific need of additional qubits $\ket{0}^{\otimes a}$. Additionally, $\hat{O}$ must have a spectral norm smaller than $1$ to ensure the existence of a block-encoding.}. The qubit state $ \ket{\psi}\ket{0}^{\otimes a}\ket{0}$ evolves through quantum circuit presented in Figure \ref{fig: Hadamard test circuit} as:
\begin{equation}
    \begin{split}
        \ket{\psi}\ket{0}^{\otimes a}\ket{0}&\xrightarrow{\hat{I}_n\otimes\hat{I}_a\otimes\hat{H}}  \ket{\psi}\ket{0}^{\otimes a}\frac{\ket{0}+\ket{1}}{\sqrt{2}} \\
        &\xrightarrow{C^1(\hat{U}_{\hat{O}})}
        \frac{1}{\sqrt{2}}(\ket{\psi}\ket{0}^{\otimes a}\ket{0}+\hat{U}_{\hat{O}}\ket{\psi}\ket{0}^{\otimes a}\ket{1}) \\ 
        &\xrightarrow{\hat{I}_n\otimes\hat{I}_a\otimes\hat{H}} 
        \frac{\hat{I}+\hat{U}_{\hat{O}}}{2}\ket{\psi}\ket{0}^{\otimes a}\ket{0}+\frac{\hat{I}-\hat{U}_{\hat{O}}}{2}\ket{\psi}\ket{0}^{\otimes a}\ket{1}.
    \end{split}
\end{equation}
The probabilities $P(0)$ and $P(1)$ of measuring the ancilla qubit in state $\ket{0}$ and $\ket{1}$ are 
\begin{equation}
    \begin{split}
        P(0)&=\|\frac{\hat{I}+\hat{U}_{\hat{O}}}{2}\ket{\psi}\ket{0}^{\otimes a}\|_2^2=\frac{1+\Re(\bra{\psi}\bra{0}^{\otimes a}\hat{U}_{\hat{O}} \ket{\psi}\ket{0}^{\otimes a})}{2}=\frac{1+\Re(\bra{\psi}\hat{O}\ket{\psi})}{2},
        \\P(1)&=\frac{1-\Re(\bra{\psi}\hat{O}\ket{\psi})}{2}.
    \end{split}
\end{equation}
By performing several times the Hadamard test, one can estimate $P(0)$ and $P(1)$ and deduce the value $\Re(\bra{\psi}\hat{O}\ket{\psi})=P(0)-P(1)$. The Hoeffding inequality applied to the Bernoulli random variables associated with the measurement results implies that $O(\log(1/\delta)/\epsilon^2)$ measurements gives an estimation of $P(0)-P(1)$ up to an error $\epsilon>0$ and a success probability at least $1-\delta$ \cite{shang2024estimating}.

Similarly, it is possible to measure the imaginary part $\Im(\bra{\psi}\hat{O}\ket{\psi})$ by adding a  phase gate $\hat{P}=\ket{0}\bra{0}+i\ket{1}\bra{1}$ on the ancilla qubit between the two Hadamard gates\footnote{Phase gates commute with the control operation.}.

\subsubsection{SWAP test}

\begin{figure}[ht]
    \centering
\begin{quantikz}
  \ket{0}\text{ } &\gate{\hat{H}} 
&\ctrl{2} & \gate{\hat{H}}&  \meter{} 
\\
&\ket{\psi}\text{ } & \swap{1} & \qw
\\ &\ket{\phi}\text{ } & \targX{} &\qw
\end{quantikz} 
\caption{Quantum circuit associated with the Swap test to estimate $\Re(\braket{\phi|\psi})$. One can estimate $\Im(\braket{\phi|\psi})$ by adding a phase gate $\hat{P}=\ket{0}\bra{0}+i\ket{1}\bra{1}$ on the ancilla qubit after the first Hadamard gate. The controlled SWAP gate corresponds to a product $n$ controlled-two-qubit SWAP gates acting as $C^1(\widehat{SWAP})\ket{\psi}\ket{\phi}\ket{0}=\ket{\psi}\ket{\phi}\ket{0}$ and $C^1(\widehat{SWAP})\ket{\psi}\ket{\phi}\ket{1}=\ket{\phi}\ket{\psi}\ket{1}$.  }
\label{fig: Swap test}
\end{figure}
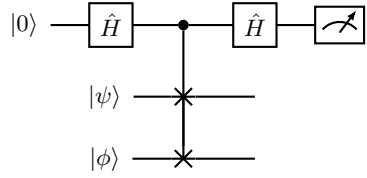

The SWAP test presented in Figure \ref{fig: Swap test} is a simple measurement protocol to estimate the overlap between two $n$-qubit quantum states $\ket{\psi}$ and $\ket{\phi}$ that are prepared on distinct registers. The control SWAP gate corresponds to $n$-parallel two-qubit SWAP gates acting on each pair of qubits such that $C^1(\widehat{SWAP})\ket{\psi}\ket{\phi}\ket{0}=\ket{\psi}\ket{\phi}\ket{0}$ and $C^1(\widehat{SWAP})\ket{\psi}\ket{\phi}\ket{1}=\ket{\phi}\ket{\psi}\ket{1}$. The quantum circuit of the SWAP test acts on $2n+1$ qubits as: 

\begin{equation}
    \begin{split}
        \ket{\psi}\ket{\phi}\ket{0} &\xrightarrow{\hat{I}_n\otimes\hat{I}_n\otimes\hat{H}} \ket{\psi}\ket{\phi}\frac{\ket{0}+\ket{1}}{\sqrt{2}} \\
        &\xrightarrow{C^1(\text{SWAP})} \frac{1}{\sqrt{2}}(\ket{\psi}\ket{\phi}\ket{0}+\ket{\phi}\ket{\psi}\ket{1})\\
        &\xrightarrow{\hat{I}_n\otimes\hat{I}_n\otimes\hat{H}}\frac{\ket{\psi}\ket{\phi}+\ket{\phi}\ket{\psi}}{2}\ket{0}+\frac{\ket{\psi}\ket{\phi}-\ket{\phi}\ket{\psi}}{2}\ket{1}.
    \end{split}
\end{equation}
The probabilities of measuring the ancilla in state $\ket{0}$ and $\ket{1}$ are:
\begin{equation}
    \begin{split}
        P(0)&=\frac{1+\Re(\braket{\phi|\psi})}{2},\\
        P(1)&=\frac{1-\Re(\braket{\phi|\psi})}{2}.
    \end{split}
\end{equation}
and the difference $P(0)-P(1)$ gives the value $\Re(\braket{\phi|\psi})$. Similarly to the Hadamard test, one can implement a phase gate $\hat{P}=\ket{0}\bra{0}+i\ket{1}\bra{1}$ on the ancilla qubit, between the two Hadamard gates, to infer $\Im(\braket{\phi|\psi})$ with the same protocol. The complexity to estimate $P(0)-P(1)$ up to an error $\epsilon>0$ with a success probability at least $1-\delta$ scales as $O(\log(1/\delta)/\epsilon^2)$.

The Swap test can also estimate average values of observable $\hat{O}$ by considering the two quantum states $\ket{\psi}\ket{0}^{\otimes a}$ and $\ket{\phi}=\hat{U}_{\hat{O}}\ket{\psi}\ket{0}^{\otimes a}$, where $\hat{U}_{\hat{O}}$ is a block-encoding of $\hat{O}$ using $a$ ancilla qubits. The depth associated with the SWAP test is smaller than the depth of the Hadamard test thanks to the fact that $\hat{U}_{\hat{O}}$ is not controlled by an ancilla qubit, thus simplifying its implementation. However, the SWAP test requires more qubits compared to the Hadamard test.

\subsubsection{Quantum amplitude estimation}

Quantum amplitude estimation (QAE) protocols provide an estimation of real-valued overlap between two qubit states $\ket{\phi}$ and $\ket{\psi}=a\ket{\phi}+\sqrt{1-a^2}\ket{\phi_\perp}$, where it is assumed that $a=\braket{\psi|\phi}\in[0,1]$ and $\braket{\phi|\phi_\perp}=0$. The main idea of QAE is to construct an operator $\hat{Q}$ whose eigenvalues are related to $e^{\pm i \theta}$, where $\cos(\theta)=a$, and to use the QPE protocol to estimate $\theta$. 

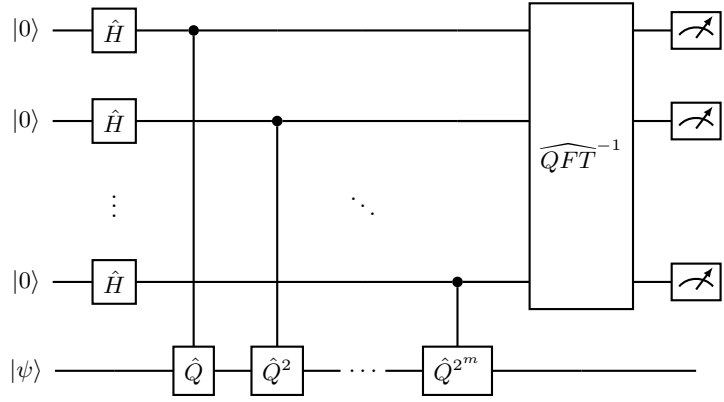
\begin{figure}[h]
    \centering
\begin{quantikz}
 &  \ket{0}\text{ }& \gate{\hat{H}}  
&\ctrl{4} &\qw&\qw&\qw& \gate[4,nwires={3}]{\widehat{QFT}^{-1}} &\meter{} 
\\
&\ket{0}\text{ } & \gate{\hat{H}} & \qw 
& \ctrl{3} & \qw &\qw&\qw&\meter{}
\\ && \vdots & & &\ddots&&&
\\
&\ket{0}\text{ } & \gate{\hat{H}} & \qw&\qw & \qw
& \ctrl{1} & \qw  &\meter{}
\\ &\ket{\psi}\text{ }&\qw&\gate{\hat{Q}} &\gate{\hat{Q}^2} &\qw \ \ldots\ & \gate{\hat{Q}^{2^m}}  &\qw&\qw&&
\end{quantikz}
\caption{Canonical quantum amplitude estimation where $\hat{Q}=\hat{U}_{\psi}\hat{S}_0\hat{U}_{\psi}^{\dagger}\hat{U}_{\phi}\hat{S}_0\hat{U}_{\phi}^{\dagger}=(\hat{I}_n-2\ket{\psi}\bra{\psi})(\hat{I}_n-2\ket{\phi}\bra{\phi})$.}
\label{fig: quantum amplitude estimation}
\end{figure}

The operator $\hat{Q}$ can be designed from the quantum state preparation unitaries $\hat{U}_\psi\ket{0}^n=\ket{\psi}$, $\hat{U}_\phi\ket{0}^n=\ket{\phi}$ \footnote{Unitary quantum state preparation is a particular type of quantum state preparation protocol that do not leverage measurements \cite{mottonen2004transformation,moosa2023linear}.} and operator $\hat{S}_0=\hat{I}_n-2\ket{0}^{\otimes n}\bra{0}^{\otimes n}$ as

\begin{equation}
    \hat{Q}=\hat{U}_{\psi}\hat{S}_0\hat{U}_{\psi}^{\dagger}\hat{U}_{\phi}\hat{S}_0\hat{U}_{\phi}^{\dagger}=(\hat{I}_n-2\ket{\psi}\bra{\psi})(\hat{I}_n-2\ket{\phi}\bra{\phi}).
\end{equation}
Note that $\hat{S}_0$ is a multi-controlled Z-Pauli gate controlled by the state $\ket{0}^{\otimes n}$. One can show that the operator $\hat{Q}$ preserves the subspace generated by $\ket{\psi}$ and $\ket{\phi}$. By denoting $a=\cos(\theta)$, $\hat{Q}$ acts on $\ket{\phi}$ and $\ket{\phi_\perp}$ as a rotation $\hat{R}_Y$ of angle $2\theta$:
\begin{equation}
    \begin{split}
        \hat{Q}\ket{\phi}&=(2a^2-1)\ket{\phi}+2a\sqrt{1-a^2}\ket{\phi_{\perp}}=\cos(2\theta)\ket{\phi}+\sin(2\theta)\ket{\phi_{\perp}},\\
        \hat{Q}\ket{\phi_\perp}&=-2a\sqrt{1-a^2}\ket{\phi}+(2a^2-1)\ket{\phi_\perp}=-\sin(2\theta)\ket{\phi}+\cos(2\theta)\ket{\phi_\perp}.
    \end{split}
\end{equation}
In particular, $\hat{Q}^j$ applied on $\ket{\psi}$ results in the qubit state:
\begin{equation}
    \hat{Q}^j\ket{\psi}=\cos((2j+1)\theta)\ket{\phi}+\sin((2j+1)\theta)\ket{\phi_\perp}.
\end{equation}
The $(n+m)$-qubit state of the QAE protocol, presented in Figure \ref{fig: quantum amplitude estimation}, evolves as:
\begin{equation}
    \begin{split}
    \ket{\psi}\ket{0}^{\otimes m} &\xrightarrow{\hat{I}_n\otimes \hat{H}^{\otimes m}}\frac{1}{\sqrt{2^m}}\sum_{j=0}^{2^m-1}\ket{\psi}\ket{j}\\
    & \xrightarrow{\prod_{k=0}^{m-1} C^1_k(\hat{Q}^{2^k})}\frac{1}{\sqrt{2^m}}\sum_{j=0}^{2^m-1}(\cos((2j+1)\theta)\ket{\phi}+\sin((2j+1)\theta)\ket{\phi_\perp})\ket{j}\\
    &=\frac{1}{\sqrt{2^m}}\sum_{j=0}^{2^m-1}\frac{1}{\sqrt{2}}(e^{2ij\theta}\ket{\phi_1}+e^{-2ij\theta}\ket{\phi_2})\ket{j}\\
    & \xrightarrow{\widehat{QFT}^{-1}}
    \frac{1}{\sqrt{2}}\sum_{k=0}^{2^m-1}\frac{1}{2^m}\sum_{j=0}^{2^m-1}\big (e^{-2i\pi kj/M}e^{ij2\theta}\ket{\phi_1}+e^{-2i\pi kj/M}e^{ij(2\pi-2\theta)}\ket{\phi_2} \big )\ket{k},
    \end{split}
\end{equation}
where $\ket{\phi_1}=e^{i\theta}\frac{\ket{\phi}-i\ket{\phi_\perp}}{\sqrt{2}}$ and $\ket{\phi_2}=e^{-i\theta}\frac{\ket{\phi}+i\ket{\phi_\perp}}{\sqrt{2}}$.
Then, by denoting $2\theta=2\pi(\tilde{\theta}/2^m+\delta)$ where $\tilde{\theta}$ is the best approximation of $\theta/\pi$ up to the $m$-th digit, one gets an estimation of $\tilde{\theta}$ or $2^m-\tilde{\theta}$, which leads to the same overlap estimation $\tilde{a}=|\cos(\pi\tilde{\theta}/2^m)|=|\cos(\pi(2^m-\tilde{\theta})/2^m)|$ with a probability of success $P\ge 4/\pi^2$. Additional details can be found in the pioneering work \textit{Quantum amplitude amplification and estimation} published in 2000 by G.Brassard et al. \cite{brassard2000quantum}. The QAE protocol provides an estimation of the real-valued overlap of qubit states with $O(1/\epsilon)$ $\hat{Q}$ operators and a constant probability of success. The probability of success can be improved up to $1-\upsilon$, where $\upsilon>0$, with a complexity scaling as $O(\log(1/\upsilon)/\epsilon)$ \cite{shang2024estimating}.

Different versions of the QAE protocol have been recently developed, removing the use of an inverse QFT on the ancilla qubits \cite{grinko2021iterative,rall2023amplitude,giurgica2022low,shang2024estimating}. These protocols have shallower depth and achieve similar scalings $O(\log(1/\upsilon)/\epsilon)$ as the canonical QAE .

\section{Proofs}
\label{appendix: proofs}
For the following proofs, we consider the space $V=[0,1]^d$ and a discretization $V_{\vec{N}}=\prod_{j=1}^d\{0,1/N_j,...,(N_j-1)/N_j\}$ with $\vec{N}=(N_1,...,N_d)$ and $N_j$ equidistant points for the $j$-th axis. We define the 2-norm on $V_{\vec{N}}$ associated with a function $g$ defined on $V$ as:
\begin{equation}
    ||g||_{2,N}=\sqrt{\sum_{\vec{X}\in V_{\vec{N}}}|g(\vec{X})|^2}.
\end{equation}

In particular, when $g\in L^2(V)$, the trapezoidal rule implies that  $||g||_{2,N}$ is an approximation of the $L^2$ norm of $g$ as:
\begin{equation}
    \frac{1}{\sqrt{N}}||g||_{2,N} \xrightarrow{N_1,...,N_d \rightarrow +\infty} \sqrt{\int_{V} |g|^2}=||g||_{L^2},
\end{equation}
with $N=N_1\times...\times N_d$

Notice that the $L^2$ norm of any derivative of a solution of the transport equation Eq.(\ref{Transport equation}) is constant over time: if $f$ is a $p$-differentiable solution, then any derivative of $f$ is solution of the same PDE with the initial condition given by the corresponding derivative of $f_0$. Secondly, notice that $f^2$ or any power of any derivative of $f$ is also solution of the same PDE with the corresponding initial condition. Then, one can deduce with the divergence theorem that the $L^2$-norm of any derivative of $f$ is conserved and equal to the $L^2$ norm of the corresponding derivative of $f_0$.

\subsubsection{Proof of Lemma \ref{lemma discrete derivative operator}}
\label{appendix: proof of lemma finite difference}

Consider the discrete derivative operator $\hat{D}_j$ of order $2p$ defined Eq.(\ref{2p discrete derivative operator}) acting on a vector encoding a $(2p+1)$ continuously differentiable function $g$ : 
\begin{equation}
\ket{g}=\frac{1}{||g||_{2,N}}\sum_{\vec{X}\in V_{\vec{N}}}g(\vec{X})\ket{\vec{X}},
\end{equation}
where $||g||_{2,N}$ is the normalization factor of $\ket{g}$.

According to Taylor formula Eq.(\ref{Taylor formula}), the action of $\hat{D}_j$ on $\ket{g}$ gives the central finite difference formula approximating the first order derivative of $g$ with an accuracy of order $2p$:

\begin{equation}
    i\hat{D}_j\ket{g}=\ket{\partial_{x_j}g}+\ket{R_j},
\end{equation}
where  $\ket{\partial_{x_j}g}=\frac{1}{||g||_{2,N}}\sum_{\vec{X}\in V_{\vec{N}}}\partial_{x_j}g(\vec{X})\ket{\vec{X}}$ and 

\begin{equation}
    \ket{R_j}=\frac{(\Delta x_j)^{2p}}{||g||_{2,N}}\sum_{\vec{X}\in V_{\vec{N}}}\sum_{k=-p}^p \frac{a_k k^{2p+1}}{(2p)!}\int_0^{1}\partial_{x_j}^{(2p+1)}g(X_1,...,X_j-sk\Delta x_j ,...,X_d)(1-s)^{2p}ds\ket{\vec{X}}.
\end{equation}

Using the Cauchy-Schwarz inequality, one can show that:

\begin{equation}
    ||\ket{R_j}||_{2,N} \leq (\Delta x_j)^{2p}\tilde{C_p}\frac{\sqrt{\sup_{s\in[0,1[}||r_s||_{2,N}^2}}{||g||_{2,N}},
\label{inequality_on_discretization}
\end{equation}
with $\tilde{C}_p=\sqrt{\sum_{k=1}^p(a_k k^{2p+1}/(2p)!)^2}$  and $r_s$ being a continuous function of $s$ defined as $r_s(x_1,...,x_d)=\partial_{x_j}^{(2p+1)}g(x_1,...,x_j+s,...,x_d)$.

The quantity $\frac{\sup_{u\in[0,1[}||r_u||_{2,N}^2}{||g||_{2,N}^2}$ tends to $\frac{||\partial_{x_j}^{(2p+1)}g||_{L^2}^2}{||g||_{L^2}^2}$ in the continuous limit $N_1,...,N_d \rightarrow +\infty$ and can therefore be bounded by a quantity $\tilde{K}_j$ independent of the discretization, depending only on $g$ and its $(2p+1)$ $j$-th derivative. Finally, one has: 
\begin{equation}
    ||\ket{R_j}||_{2,N} \leq K_j (\Delta x_j)^{2p},
\label{dicretized error kj}
\end{equation}

with $K_j=\tilde{C_p}\tilde{K}_j $ a constant independent of the discretization which depends only on $p$, the $L^2$ norm of $g$ and the $(2p+1)-j$-th derivative of $g$.

\subsubsection{Proof of Theorem \ref{thm_discretization_error}}
\label{appendix: proof of thm discretization error}

The transport equation Eq.(\ref{Transport equation}) can be recast into a form where the discrete Hamiltonian $\hat{H}(t)=\sum_{j=1}^d \hat{c}_j(t)\hat{D}_j$ appear:
\begin{equation}
   \partial_t \ket{f}_t=-i\hat{H}(t)\ket{f}_t+\ket{r(t)},
\label{eq: proof thm2, evolution}
\end{equation}
with $\ket{r(t)}=\sum_{j=1}^d \hat{c}_j(t)(i\hat{D}_j\ket{f}_t-\ket{\partial_{x_j}f}_t)$ and $\ket{\partial_{x_j}f}_t=\frac{1}{||f_0||_{2,N}}\sum_{\vec{X}\in V_{\vec{N}}}\partial_{x_j}f(\vec{X},t)\ket{\vec{X}}$. The variation of parameter formula represents the solution of this equation as :
\begin{equation}
    \ket{f}_t=\hat{U}(t,0)\ket{f_0}+\int_0^t\hat{U}(t,s)\ket{r(s)}ds,
\label{eq: proof thm2, variation parameter formula}
\end{equation}
where $\hat{U}(t,s)$ is the unitary evolution operator associated with equation Eq.(\ref{ODE})n which solves the following differential equation:
\begin{equation}
    \partial_t\hat{U}(t,s)=-i\hat{H}(t)\hat{U}(t,s),\text{      } \hat{U}(s,s)=\hat{I}.
\label{eq: proof thm2, unitary evolution}
\end{equation}
Noting that $\hat{U}(t,0)\ket{f_0}$ is simply the solution $\ket{\tilde{f}}$ of the equation Eq\ref{ODE}, one can represent the difference between the two vectors as the integral over time of the vector $\hat{U}(t,s)\ket{r(s)}$:
\begin{equation}
    \ket{f}_t-\ket{\tilde{f}}_t=\int_0^t\hat{U}(t,s)\ket{r(s)}ds.
\label{eq: proof thm2, variation parameter formula 2}
\end{equation}

In terms of vector $2$-norm, the error is bounded by the time $t$ and the maximum over $s\in [0,t]$ of the vector norm of $\ket{r(s)}$ which, using Lemma \ref{lemma discrete derivative operator}, gives:
\begin{equation}
    ||\ket{f}_t-\ket{\tilde{f}}_t||_{2,N}\leq t \sum_{j=1}^d K_j ||c_j||_\infty (\Delta x_j)^{2p} .
\end{equation}
By defining $K=\max_{j=1,...,d}K_j$ one concludes the proof.

\subsubsection{Proofs of Theorem \ref{thm_product_formula_error_operator_norm} and Theorem \ref{thm_product_formula_error}}
\label{appendix: proof of trotterization error}

In this appendix, the operator norm inequality and the vector norm inequality are derived. The first steps are similar for the two inequalities. Then, additional computations are required to derive the tighter vector norm inequality.

In order to approximate the precise unitary progression within a timeframe T, we break down the time span $[0, T]$ into $L$ equal segments and apply Trotter discretization to each segment. As both continuous and discretized evolutions are unitary, the overall error is essentially the sum of individual errors at each step.

\begin{equation}
\begin{split}
     & ||\ket{\tilde{f}_{\alpha}}_T-\ket{\tilde{f}}_T||_{2,N}=||\prod_{l=1}^L\hat{U}_{\alpha}(\frac{lT}{L},\frac{(l-1)T}{L})\ket{f_0}-\hat{U}(T,0)\ket{f_0}||_{2,N}\\ = & ||\prod_{l=1}^L\hat{U}_{\alpha}(\frac{lT}{L},\frac{(l-1)T}{L})\ket{f_0}-\prod_{l=1}^L\hat{U}(\frac{lT}{L},\frac{(l-1)T}{L})\ket{f_0}||_{2,N} \\ \leq & \sum_{k=1}^L ||(\prod_{l=k+1}^L\hat{U}_{\alpha}(\frac{lT}{L},\frac{(l-1)T}{L}))(\prod_{l=1}^k \hat{U}(\frac{lT}{L},\frac{(l-1)T}{L}))\ket{f_0} - (\prod_{l=k}^L\hat{U}_{\alpha}(\frac{lT}{L},\frac{(l-1)T}{L}))(\prod_{l=1}^{k-1} \hat{U}(\frac{lT}{L},\frac{(l-1)T}{L}))\ket{f_0}||_{2,N}   \\ \leq & \sum_{k=1}^L ||(\prod_{l=1}^k \hat{U}(\frac{lT}{L},\frac{(l-1)T}{L}))\ket{f_0} - (\hat{U}_{\alpha}(\frac{kT}{L},\frac{(k-1)T}{L}))(\prod_{l=1}^{k-1} \hat{U}(\frac{lT}{L},\frac{(l-1)T}{L}))\ket{f_0}||_{2,N}\\ = & \sum_{l=1}^L||(\hat{U}_{\alpha}(\frac{lT}{L},\frac{(l-1)T}{L})-\hat{U}(\frac{lT}{L},\frac{(l-1)T}{L}))\ket{\tilde{f}}_{\frac{(l-1)T}{L}}||_{2,N}.
\end{split}
\end{equation}

Now, we bound each difference $||(\hat{U}_{\alpha}(\frac{lT}{L},\frac{(l-1)T}{L})-\hat{U}(\frac{lT}{L},\frac{(l-1)T}{L}))\ket{\tilde{f}}_{\frac{(l-1)T}{L}}||_{2,N}$ by using the variation of parameter formula similarly to Equations \ref{eq: proof thm2, evolution}, \ref{eq: proof thm2, variation parameter formula}, \ref{eq: proof thm2, unitary evolution} and \ref{eq: proof thm2, variation parameter formula 2}:
\begin{equation}
    (\hat{U}_{\alpha}(t+h,t)-\hat{U}(t+h,t))\ket{\tilde{f}}_{t}=\int_0^h\hat{U}(t+h,t+s)\ket{r_{\alpha}'(t+s,t)}ds,
\end{equation}
with 

\begin{equation}
\begin{split}
    \ket{r_{g}'(t+s,t)} &= i\sum_{j=1}^{d-1} \big[\hat{c}_j(t+s)\hat{D}_j,(\overleftarrow{\prod}_{k=j+1}^d\hat{U}_{k,g}(t+s,t))\big]\overleftarrow{\prod}_{k'=1}^j\hat{U}_{k',g}(t+s,t)\ket{\tilde{f}}_t ,\\
   \ket{r_{s}'(t+s,t)} &= i\sum_{j=1}^{d-1} \big[\hat{c}_j(t+s)\hat{D}_j,(\overleftarrow{\prod}_{k=j+1}^d\hat{U}_{k,s}(t+s,t))\big]\overleftarrow{\prod}_{k'=1}^j\hat{U}_{k',s}(t+s,t)\ket{\tilde{f}}_t \\
    &-is\sum_{j=1}^{d}(\overleftarrow{\prod}_{k'=j+1}^d\hat{U}_{k',s}(t+s,t))\partial_t\hat{c_j}(t+h)\hat{D}_j(\overleftarrow{\prod}_{k'=1}^j\hat{U}_{k',s}(t+s,t))\ket{\tilde{f}}_t.
\end{split}
\label{eq:r terms}
\end{equation}

Now, we focus on bounding the $2$-norm of $\ket{r_{\alpha}'(t+s,t)}$ by a term proportionate to $s$, in order to get an overall error proportionates to $Lh^2=T^2/L$, where $L$ is the number of time step. 

To get a second integral over time, it is necessary to apply the Taylor theorem on the product of unitaries:
\begin{equation}  
\begin{split}
\overleftarrow{\prod}_{k=j+1}^d\hat{U}_{k,g}(t+s,t)&=\hat{I}+\int_0^sds' (\partial_s(\overleftarrow{\prod}_{k=j+1}^d\hat{U}_{k,g}))(s')\\&=\hat{I}+\int_0^s ds'\sum_{m=j+1}^d(\overleftarrow{\prod}_{k=m+1}^d\hat{U}_{k,g}(t+s',t))\hat{c}_m(t+s')\hat{D}_m(\overleftarrow{\prod}_{k=j+1}^m\hat{U}_{k,g}(t+s',t)) ,\\
\overleftarrow{\prod}_{k=j+1}^d\hat{U}_{k,s}(t+s,t)&=\hat{I}+\int_0^s ds'\sum_{m=j+1}^d(\overleftarrow{\prod}_{k=m+1}^d\hat{U}_{k,s}(t+s',t))(\hat{c}_m(t+s')+s'\partial_t \hat{c}_m(t+s'))\hat{D}_m(\overleftarrow{\prod}_{k=j+1}^m\hat{U}_{k,s}(t+s',t)).
\end{split}
\end{equation}

The commutator with the identity operator vanishes, leading to the following formula:

\begin{equation}
\begin{split}
   & \ket{r_{g}'(t+s,t)} \\ &=\int_0^s ds' \sum_{j=1}^{d-1} \sum_{m=j+1}^d \big[\hat{c}_j(t+s)\hat{D}_j,(\overleftarrow{\prod}_{k=m+1}^d\hat{U}_{k,g}(t+s',t))\hat{c}_m(t+s')\hat{D}_m(\overleftarrow{\prod}_{k=j+1}^m\hat{U}_{k,g}(t+s',t))\big]\overleftarrow{\prod}_{k=1}^j\hat{U}_{k,g}(t+s,t)\ket{\tilde{f}}_t ,\\
   & \ket{r_{s}'(t+s,t)} \\ &=\int_0^s ds' \sum_{j=1}^{d-1} \sum_{m=j+1}^d \big[\hat{c}_j(t+s)\hat{D}_j,(\overleftarrow{\prod}_{k=m+1}^d\hat{U}_{k,s}(t+s',t))\hat{c}_m(t+s')\hat{D}_m(\overleftarrow{\prod}_{k=j+1}^m\hat{U}_{k,s}(t+s',t))\big]\overleftarrow{\prod}_{k=1}^j\hat{U}_{k,s}(t+s,t)\ket{\tilde{f}}_t \\
   &+\int_0^s ds' \sum_{j=1}^{d-1} \sum_{m=j+1}^d \big[\hat{c}_j(t+s)\hat{D}_j,(\overleftarrow{\prod}_{k=m+1}^d\hat{U}_{k,s}(t+s',t))s'\partial_t\hat{c}_m(t+s')\hat{D}_m(\overleftarrow{\prod}_{k=j+1}^m\hat{U}_{k,s}(t+s',t))\big]\overleftarrow{\prod}_{k=1}^j\hat{U}_{k,s}(t+s,t)\ket{\tilde{f}}_t \\
   &-is\sum_{j=1}^{d}(\overleftarrow{\prod}_{k'=j+1}^d\hat{U}_{k',s}(t+s,t))\partial_t\hat{c_j}(t+h)\hat{D}_j(\overleftarrow{\prod}_{k'=1}^j\hat{U}_{k',s}(t+s,t))\ket{\tilde{f}}_t.
\end{split}
\end{equation}

The operator norm inequality given in Theorem \ref{thm_product_formula_error_operator_norm} can be deduced from these formula by using the norm inequality $\|\hat{A}\ket{g}\|_{2,N} \leq \|\hat{A}\|_2 \|\ket{g}\|_{2,N}$. In particular, if $\hat{A}$ is unitary $\|\hat{A}\|_2=1$ and if $\ket{g}$ is normalized $\|\ket{g}\|_{2,N}=1$. By defining $||c_j||_{\infty}=\sup_{\vec{x}\in [0,1[^d,t\in[0,T]}|c_j(\vec{x},t)|$ and noticing $\|\hat{c}_j\|_2\leq||c_j||_{\infty}$, one can conclude: 

\begin{equation}
\begin{split}
    ||\ket{\tilde{f}_{g}}_T-\ket{\tilde{f}}_T||_{2,N}& \leq  \frac{T^2}{L} \sum_{j=1}^{d-1}\sum_{m=j+1}^d \|c_j\|_{\infty} \| \|c_m\|_{\infty} \|\hat{D}_j\|_2\|\hat{D}_m\|_2 ,   
\\
    ||\ket{\tilde{f}_{s}}_T-\ket{\tilde{f}}_T||_{2,N}&\leq  \frac{T^2}{L}\bigg(\sum_{j=1}^{d-1}\sum_{m=j+1}^d \|c_j\|_{\infty} \| \|c_m\|_{\infty} \|\hat{D}_j\|_2\|\hat{D}_m\|_2 +\frac{1}{2}\sum_{j=1}^d\|\partial_tc_j\|_{\infty} \|\hat{D}_j\|_2\bigg) \\&+\frac{T^3}{L^2} \frac{1}{3} \sum_{j=1}^{d-1}\sum_{m=j+1}^d \|c_j\|_{\infty}\|\partial_t c_m\|_{\infty} \|\hat{D}_j\|_2\|\hat{D}_m\|_2 .
\end{split}
\label{Eq: Scaling_proof}
\end{equation}

Additional computations are required to achieve a bound that do not scale  with the spectral norm of the derivative operators, i.e., with $1/\Delta x_j^2$. In the following, we derive the vector norm inequality for the general product formula $\alpha=g$. The analysis of the standard product formula is similar. 

From this step of the proof, reference \cite{An2021} which deals with a Hamiltonian $H=H_1+H_2$, used a Gronwall inequality and the fact that one Hermitian operator among $H_1,H_2$ is bounded to derive a vector norm inequality for terms of the form $||\hat{H}_1 (\prod_k \hat{U}_{k})\hat{H}_2 (\prod_k \hat{U}_{k})\ket{\tilde{f}}_t ||_{2,N}$. This approach is not applicable here since all the summands in the Hamiltonian are asymptotically unbounded due to the derivative operators $\|\hat{D}_j\|_2=\Theta(2^n)$. One way to circumvent this issue is to use the definition of the discrete derivative operator via Lemma \ref{lemma discrete derivative operator}: a discrete derivative operator applied on a vector depending on a differentiable function can be approximated by the vector encoding the derivative of the function. Therefore, we need to introduce $\ket{f}_t$, the real space encoding of the solution of the transport equation at time $t$: $\ket{\tilde{f}_t}=\ket{\tilde{f}}_t-\ket{f}_t+\ket{f}_t$. 

\begin{equation}
\begin{split}
   & \int_0^hds \|\ket{r_{g}'(t+s,t)} \|_{2,N} \\ & \leq \int_0^hds\int_0^s ds' \sum_{j=1}^{d-1} \sum_{m=j+1}^d \biggl(  \bigg\|\big[\hat{c}_j(t+s)\hat{D}_j,(\overleftarrow{\prod}_{k=m+1}^d\hat{U}_{k,g}(t+s',t))\hat{c}_m(t+s')\hat{D}_m(\overleftarrow{\prod}_{k=j+1}^m\hat{U}_{k,g}(t+s',t))\big]\bigg \|_2 \bigg\|\ket{\tilde{f}}_t-\ket{f}_t \bigg\|_{2,N}    \\ & +\bigg\| \big[\hat{c}_j(t+s)\hat{D}_j,(\overleftarrow{\prod}_{k=m+1}^d\hat{U}_{k,g}(t+s',t))\hat{c}_m(t+s')\hat{D}_m(\overleftarrow{\prod}_{k=j+1}^m\hat{U}_{k,g}(t+s',t))\big]\overleftarrow{\prod}_{k=1}^j\hat{U}_{k,g}(t+s,t)\ket{f}_t \bigg \|_{2,N} \biggl).
\end{split}
\label{eq:inequality 1.65}
\end{equation}

The first term can be bounded using the discretization error given by Theorem \ref{thm_discretization_error} and by the maximum value over space and time of the $c_j$ function and the operator norm of $\hat{D}_j$: 

\begin{equation}
\begin{split}
  &\int_0^hds\int_0^s ds' \sum_{j=1}^{d-1} \sum_{m=j+1}^d \bigg \|\big[\hat{c}_j(t+s)\hat{D}_j,(\overleftarrow{\prod}_{k=m+1}^d\hat{U}_{k,g}(t+s',t))\hat{c}_m(t+s')\hat{D}_m(\overleftarrow{\prod}_{k=j+1}^m\hat{U}_{k,g}(t+s',t))\big] \bigg\|_2 \bigg\|\ket{\tilde{f}}_t-\ket{f}_t \bigg\|_{2,N}  \\ &\leq\tilde{K} d(d-1)h^2T\max_j \|c_j\|_\infty^3 \max_{j,j',m}\frac{(\Delta x_j)^{2p}}{\Delta x_{j'} \Delta x_m} ,
\end{split}
\label{eq: termone}
\end{equation}
with $\tilde{K}=2C_p^2K$, $C_p=\sum_{q=0}^p|a_q|$, the $a_q$ being the coefficients of the central finite difference and $K=\max_jK_j$ is given in Equation \ref{dicretized error kj}. The term $\|\ket{\tilde{f}}_t-\ket{f}_t \|_{2,N}$ induce a factor $(\Delta x_j)^{2p}$ that counterbalance the $1/\Delta x_{j'}$ and $1/(\Delta x_m)$ for $p\ge 2$. In other words, when the discretization error $\|\ket{\tilde{f}}_t-\ket{f}_t \|_{2,N}$ is $\epsilon$-small and after $L$ time-steps, the term Eq.(\ref{eq: termone}) scales as $O(L\frac{T^2}{L^2\Delta x^2}\epsilon)=O(\epsilon^{2-1/(2p)})$

The second term in inequality \ref{eq:inequality 1.65} can be bounded using the Taylor formula for each product of unitaries:
\begin{equation}  
\begin{split}&\overleftarrow{\prod}_{k=m+1}^d\hat{U}_{k,g}(t+s',t)=\hat{I}+\int_0^{s'}ds_2\sum_{k=m+1}^d\hat{A}_{1,k}(t+s_2,t), \\
&\overleftarrow{\prod}_{k=j+1}^m\hat{U}_{k,g}(t+s',t)=\hat{I}+\int_0^{s'}ds_3\sum_{k=j+1}^m \hat{A}_{2,k}(t+s_3,t), \\
& \overleftarrow{\prod}_{k=1}^j\hat{U}_{k,g}(t+s,t)=\hat{I}+\int_0^{s}ds_4\sum_{k=1}^j \hat{A}_{3,k}(t+s_4,t) ,
\end{split}
\end{equation}
with 
\begin{equation}
\begin{split}
    &\hat{A}_{1,k}(t+s_2,t)=(\overleftarrow{\prod}_{k'=k+1}^d\hat{U}_{k',g}(t+s_2,t))\hat{c}_k(t+s_2)\hat{D}_k(\overleftarrow{\prod}_{k'=m+1}^k\hat{U}_{k',g}(t+s_2,t)), \\
    &\hat{A}_{2,k}(t+s_3,t)=(\overleftarrow{\prod}_{k'=k+1}^m\hat{U}_{k',g}(t+s_3,t))\hat{c}_k(t+s_3)\hat{D}_k(\overleftarrow{\prod}_{k'=j+1}^k\hat{U}_{k',g}(t+s_3,t)),
    \\
    & \hat{A}_{3,k}(t+s_4,t)=(\overleftarrow{\prod}_{k'=k+1}^j\hat{U}_{k',g}(t+s_4,t))\hat{c}_k(t+s_4)\hat{D}_k(\overleftarrow{\prod}_{k'=1}^k\hat{U}_{k',g}(t+s_4,t)).
\end{split}
\end{equation}

These three Taylor formula leads to eight different terms. The first term involves the three identity operator and can be bounded using Lemma \ref{lemma discrete derivative operator}. The other terms involve one, two or three additional integral over time, giving terms proportionate to $h^3$, $h^4$ and $h^5$ that are negligible compared to the first one. The first term is bounded by $h^2$ as

\begin{equation}
\begin{split}
    &\int_0^hds\int_0^s ds' \sum_{j=1}^{d-1} \sum_{m=j+1}^d\bigg \| \big[\hat{c}_j(t+s)\hat{D}_j,\hat{c}_m(t+s')\hat{D}_m \big]\ket{f}_t \bigg \|_{2,N} \\&\leq \frac{h^2}{2}  \sum_{j=1}^{d-1} \sum_{m=j+1}^d \bigg( \|c_j\|_{\infty} \|\partial_{x_j} c_m\|_{\infty} \frac{ \max_{t\in[0,T]}\|\partial_{x_m}f_t\|_{2,N}}{\|f_0\|_{2,N}}
     +\|c_m\|_{\infty} \|\partial_{x_m} c_j\|_{\infty} \frac{\max_{t\in[0,T]} \|\partial_{x_j}f_t\|_{2,N}}{\|f_0\|_{2,N}} \bigg)
    \\&+O(h^2\max_j\Delta x_j ^{2p}),
\end{split}
\label{eq:leading term vector norm inequality generalized formula}
\end{equation}
where $\|\partial_{x_m}f_t\|_{2,N}=\sqrt{\sum_{\vec{X}}|\partial_{x_m}f(\vec{X},t)|^2}$ and $\|f_0\|_{2,N}=\sqrt{\sum_{\vec{X}}|f_0(\vec{X})|^2}$.
Remark that the commutator form implies that the terms involving the second derivative of $f$ cancel due to the Schwartz equality $\partial_{x_j}\partial_{x_m}f=\partial_{x_m}\partial_{x_j}f$. Remark also that $\frac{ \|\partial_{x_m}f_t\|_{2,N}}{\|f_0\|_{2,N}}$ converges in the large $N$ limit toward $\frac{ \|\partial_{x_m}f_t\|_{L^2}}{\|f_0\|_{L^2}}=\frac{\|\partial_{x_m}f_0\|_{L^2}}{\|f_0\|_{L^2}}$, where $f_t=f(t,.)$ and $\|f\|_{L^2}=\sqrt{\int_{[0,1]^d}|f(x)|^2dV}$

The seven other terms can be bounded using the $\|\hat{D}_j\|\leq2C_p/\Delta x_j$ since they are proportionate to higher power of $h=T/L$ as:
\begin{equation}
\begin{split}
       &\int_0^hds\int_0^s ds' \int_0^{s'} ds_2 \sum_{j=1}^{d-1} \sum_{m=j+1}^d \sum_{k=m+1}^d\bigg \| \big[\hat{c}_j(t+s)\hat{D}_j,\hat{A}_{1,k}(t+s_2,t)\hat{c}_m(t+s')\hat{D}_m \big]\ket{f}_t \bigg \|_{2,N} \\ &\leq \frac{4}{9}h^3 d(d-1)(d-2)(C_p)^3\max_j\|c_j\|_{\infty}^3 \max_{j}\frac{1}{(\Delta x_j)^3},
       \\&\int_0^hds\int_0^s ds' \int_0^{s'} ds_3 \sum_{j=1}^{d-1} \sum_{m=j+1}^d \sum_{k'=j+1}^m
       \bigg \| \big[\hat{c}_j(t+s)\hat{D}_j,\hat{c}_m(t+s')\hat{D}_m \hat{A}_{2,k'}(t+s_3,t)\big]\ket{f}_t \bigg \|_{2,N} \\&\leq \frac{4}{9}h^3d(d^2-1)(C_p)^3\max_j\|c_j\|_{\infty}^3 \max_{j}\frac{1}{(\Delta x_j)^3},
       \\&\int_0^hds\int_0^s ds' \int_0^{s} ds_4 \sum_{j=1}^{d-1} \sum_{m=j+1}^d \sum_{k"=1}^j
       \bigg \| \big[\hat{c}_j(t+s)\hat{D}_j,\hat{c}_m(t+s')\hat{D}_m \big]\hat{A}_{3,k"}(t+s_4,t)\ket{f}_t \bigg \|_{2,N} \\ & \leq \frac{8}{9}h^3d(d^2-1)(C_p)^3\max_j\|c_j\|_{\infty}^3 \max_{j}\frac{1}{(\Delta x_j)^3},
       \\&\int_0^hds\int_0^s ds' \int_0^{s'} ds_2 \int_0^{s'} ds_3 \sum_{j=1}^{d-1} \sum_{m=j+1}^d \sum_{k=m+1}^d
       \sum_{k'=j+1}^m
       \bigg \| \big[\hat{c}_j(t+s)\hat{D}_j,\hat{A}_{1,k}(t+s_2,t)\hat{c}_m(t+s')\hat{D}_m \hat{A}_{2,k'}(t+s_3,t)\big]\ket{f}_t \bigg \|_{2,N}\\&\leq \frac{1}{9}h^4d(d^2-1)(d-2)(C_p)^4\max_j\|c_j\|_{\infty}^4 \max_{j}\frac{1}{(\Delta x_j)^4} ,
       \\&\int_0^hds\int_0^s ds' \int_0^{s'} ds_2 \int_0^{s} ds_4\sum_{j=1}^{d-1} \sum_{m=j+1}^d \sum_{k=m+1}^d \sum_{k"=1}^j
       \bigg \| \big[\hat{c}_j(t+s)\hat{D}_j,\hat{A}_{1,k}(t+s_2,t)\hat{c}_m(t+s')\hat{D}_m \big]\hat{A}_{3,k"}(t+s_4,t)\ket{f}_t \bigg \|_{2,N} \\& \leq \frac{1}{6}h^4d(d^2-1)(d-2)(C_p)^4\max_j\|c_j\|_{\infty}^4 \max_{j}\frac{1}{(\Delta x_j)^4} ,
        \\&\int_0^hds\int_0^s ds' \int_0^{s'} ds_3 \int_0^{s} ds_4\sum_{j=1}^{d-1} \sum_{m=j+1}^d \sum_{k'=j+1}^m \sum_{k"=1}^j
       \bigg \| \big[\hat{c}_j(t+s)\hat{D}_j,\hat{c}_m(t+s')\hat{D}_m \hat{A}_{2,k'}(t+s_3,t)\big]\hat{A}_{3,k"}(t+s_4,t)\ket{f}_t \bigg \|_{2,N} \\& \leq \frac{1}{6}h^4d(d^2-1)(d+2)(C_p)^4\max_j\|c_j\|_{\infty}^4 \max_{j}\frac{1}{(\Delta x_j)^4} , 
          \\&\int_0^hds\int_0^s ds' \int_0^{s'} ds_2 \int_0^{s'} ds_3 \int_0^{s} ds_4  \sum_{j=1}^{d-1} \sum_{m=j+1}^d \sum_{k=m+1}^d \sum_{k'=j+1}^m \sum_{k"=1}^j
       \bigg \| \big[\hat{c}_j(t+s)\hat{D}_j,\hat{A}_{1,k}(t+s_2,t)\hat{c}_m(t+s')\hat{D}_m\hat{A}_{2,k'}(t+s_3,t) \big]\\&\hat{A}_{3,k"}(t+s_4,t)\ket{f}_t \bigg \|_{2,N}  \leq \frac{2}{75}h^5d(d^2-1)(d^2-4)(C_p)^5\max_j\|c_j\|_{\infty}^5\max_{j}\frac{1}{(\Delta x_j)^5} .
\end{split}
\end{equation}

To conclude, after $L$ time steps the term of lowest order in $h=T/L$ is given Equation \ref{eq:leading term vector norm inequality generalized formula} and an exact inequality can be derived for all other terms. These terms are asymptotically negligible compared to the leading term: they scale as $O(\frac{T^q}{L^{q-1}}\max_j1/\Delta x^q)$ with $q\ge3$ and by choosing $L=O(T^2/\epsilon)$ and $\Delta x=O((\epsilon/T)^{1/2p})$ (see Eq.(\ref{qubit scaling}), these terms scales as $O(\epsilon^{3(1-1/(2p))-1}/T^{3(1-1/(2p))-2})$. In the worst case $q=3, p=2$, the scaling is $O(\epsilon^{5/4}/T^{1/4})$ which is negligible compared to $\epsilon$.

The proof for the standard product formula is identical and the leading order has an additional term equal to: \begin{equation}
\frac{1}{2}\sum_{j=1}^d\|\partial_tc_j\|_{\infty} \max_{t\in[0,T]}\frac{\|\partial_{x_j}f_t\|_{2,N}}{\|f_0\|_{2,N}}.
\end{equation} 

\end{document}